# Divergence-aware adaptive prediction framework for accelerating CFD simulations of unsteady flows

**Authors:** Xiangrui Zou[1*], Zhuoqun Zhao[1,2], Guillermo Barragán[1], Soledad Le Clainche[1*]

[1] ETSI Aeronautica y del Espacio, Universidad Politécnica de Madrid, Plaza Cardenal Cisneros, 3, 28040, Madrid, Spain.

[2] ETSI Telecomunicación, Universidad Politécnica de Madrid, Av. Complutense, 30, 28040, Madrid, Spain.

**Corresponding Author**:

Xiangrui Zou[1]*, Email: x.zou@upm.es, Tel/Fax: +34 600 67 89 35

ETSI Aeronautica y del Espacio, Universidad Politécnica de Madrid, Plaza Cardenal Cisneros, 3, 28040, Madrid, Spain.

Soledad Le Clainche[1]*, Email: soledad.leclainche@upm.es, Tel/Fax: +34 910 67 58 53

ETSI Aeronautica y del Espacio, Universidad Politécnica de Madrid, Plaza Cardenal Cisneros, 3, 28040, Madrid, Spain.

**Abstract**

Reliable long-horizon prediction remains a challenge for data-driven CFD surrogates, because offline-trained models accumulate autoregressive errors and lose accuracy when operating conditions change. This work develops a divergence-aware adaptive CFD-surrogate framework that couples a CFD solver with a proper orthogonal decomposition-deep learning (POD-DL) surrogate in a closed-loop workflow. CFD snapshots are compressed by POD, and a neural-network predictor advances the reduced state in time. The surrogate performs autoregressive forecasting, while its reliability is monitored online. When a prescribed update interval is reached or prediction degradation is detected, the CFD solver is automatically recalled to generate new snapshots and update the surrogate. The framework is assessed for three-dimensional flow past a circular cylinder at Re = 160–400. Baseline non-adaptive predictions exhibit progressive error growth over long forecast horizons, confirming the need for online correction. With prescribed update intervals, the adaptive framework preserves the dominant wake dynamics and reduces error growth after retraining compared with the non-adaptive model. For a representative 200-snapshot interval, the framework achieves a speed-up ratio of approximately 92 relative to CFD. An event-triggered mode is introduced using ensemble uncertainty and dynamically estimated thresholds. This mode terminates unreliable forecasts without requiring ground-truth CFD data during prediction, and the detected triggers are consistent with the onset of deterioration in the lift-coefficient evolution. Under varying inlet conditions, the framework detects regime changes, recalls CFD, and recovers reliable predictions. These results demonstrate that divergence-aware CFD-surrogate coupling provides a robust and efficient route for adaptive long-horizon flow prediction under evolving operating conditions.

## 1. Introduction

Computational fluid dynamics (CFD) simulations are widely used across fluid mechanics, combustion, meteorology and related disciplines, and they play a central role in analysis, design and optimization processes [1,2]. In general, predictive accuracy relies on resolving the relevant spatiotemporal scales of the underlying physics. However, many realistic configurations involve large geometric domains and multiscale dynamics, making high-fidelity simulations prohibitively expensive. As a result, achieving sufficiently high resolution for large-scale complex systems often requires computational resources that are impractical in routine engineering workflows [3,4]. This motivates the development of surrogate and acceleration frameworks that can replace or substantially speed up CFD while preserving accuracy [1,2].

To mitigate this computational burden, reduced-order models (ROMs) have become a standard strategy: the high-dimensional flow state is first compressed into a compact latent representation, and a learning-based predictor is then trained to advance this representation in time, providing an efficient alternative to repeated high-fidelity simulations [5]. In this context, projection-based ROMs build a low-dimensional subspace from snapshots of the full-order model (FOM) and derive reduced dynamics through Galerkin-type projection, demonstrating effectiveness in aerodynamics, flow control, and combustion applications [6–8].

In parallel, non-intrusive data-driven ROMs bypass explicit access to the governing equations and instead infer the reduced dynamics directly from data, which makes them attractive for complex solvers and industrial workflows [9,10]. Representative hybrid approaches combine modal decomposition (proper orthogonal decomposition, POD; dynamic mode decomposition, DMD) or nonlinear encoders with recurrent or convolutional neural networks to model temporal evolution, achieving substantial speedups for canonical unsteady flows while retaining competitive accuracy [11–13]. Le Clainche et al. [14–19] further developed a family of predictive hybrid ROMs that couple modal decomposition with deep learning architectures for a range of applications, including cylinder wakes, synthetic jets, two-phase flows and jet combustion.

Although ROMs and learning-based surrogates often deliver substantial speedups under offline training, their reliability may deteriorate under distribution shift and over long prediction horizons due to error accumulation [20–22]. To address this issue, one line of research seeks to improve the intrinsic predictive capability of the surrogate, whereas another develops adaptive prediction strategies that can recall the high-fidelity solver once the surrogate begins to diverge [20–22]. To enhance predictive capability, Özalp et al. [23] proposed a data-assimilation framework, DA-CAE-ESN, which combines a convolutional autoencoder (CAE), an echo state network (ESN), and an ensemble Kalman filter (EnKF) for real-time forecasting of high-dimensional spatiotemporal chaotic systems. Using the Kuramoto-Sivashinsky equation and the two-dimensional Navier-Stokes equations, they showed that the framework can maintain stable and accurate predictions under varying noise levels, sparsity, and sampling rates. Zighed et al. [24] proposed a dynamic ROM for unsteady-flow prediction. Their model combines a variational autoencoder (VAE) for nonlinear dimensionality reduction and uncertainty quantification with a Transformer-based latent-space dynamical model. Results on canonical unsteady flows demonstrate improved parametric generalization, uncertainty awareness, and adaptive sampling efficiency, highlighting the potential of confidence-aware surrogate for robust prediction and decision support.

However, as the underlying system evolves, the operating conditions may drift beyond the range represented in the training data [20–22]. This has motivated the development of adaptive learning strategies that update the reduced representation or the surrogate model when new dynamics are encountered. One important direction is based on projection-based ROMs. Terragni and Vega [25] used POD on the fly to construct bifurcation diagrams. Their method combines a standard numerical solver with a Galerkin system obtained by projecting the governing equations onto POD modes. These modes are computed and updated as the bifurcation parameter varies. The need for updating is detected online by monitoring the amplitudes of high-order modes and the consistency with an instrumental Galerkin system. However, the method relies on projected governing equations and online consistency monitoring between reduced systems. Rapún et al. [26] developed an adaptive POD-based low-dimensional modelling strategy to simulate time-dependent dynamics in nonlinear dissipative systems. Their method combines a POD-based Galerkin system with short runs of a standard numerical solver, and the switching between both models is decided on the fly using truncation-error and residual estimates. While effective, the approach requires access to intrusive reduced equations and residual evaluation during the simulation. Le Clainche et al. [27] further applied POD on the fly to accelerate long-term subsurface oil/water-flow simulations. In their method, POD modes are first calculated from a short initial run of the full model and then updated along the simulation using new snapshots computed in shorter additional runs, allowing the reduced basis to adapt to the local dynamics. Nevertheless, the reduced dynamics still depend on projection-based reduced-order formulations.

Another related direction is based on data-driven ROMs. A representative work is the equation-free multiscale paradigm [28–30], which couples an expensive fine-scale simulator with a low-dimensional coarse description. In this setting, the fine-scale model is executed for a short time window to generate fine-scale trajectories, which are then lifted into a reduced set of coarse variables. The coarse variables are subsequently advanced using inexpensive time integrators to approximate the macro-scale evolution. Beltrán et al. [27] proposed an adaptive data-driven ROM based on higher-order dynamic mode decomposition (HODMD)[31], where a numerical solver is alternated with HODMD-based extrapolation intervals to adapt the prediction to the varying dynamics along transient simulations. Yet, the method remains tied to modal extrapolation strategies with limited adaptability to strongly evolving nonlinear dynamics. More recently, several adaptive surrogate frameworks have been proposed that explicitly alternate between a computational solver and a learned predictor. For example, Kicic et al. [20] introduced an online scheme that learns an effective reduced dynamics model while delegating unexplored regimes to the high-fidelity solver; the surrogate is updated continuously through online training, and switching is triggered based on monitored predictive accuracy and uncertainty. Scherding et al. [21] developed an adaptive strategy for high-dimensional surrogate modeling that combines nonlinear dimensionality reduction with clustering and local regression, together with active learning to mitigate extrapolation; the method was demonstrated on chemically nonequilibrium hypersonic flows with notable cost reduction. Abadia-Heredia et al. [22] proposed a data-driven adaptive autoregressive forecasting framework based on generated datasets, which improves long-horizon stability by updating the predictor when degradation is detected by comparing with ground truth. As a result, the development of adaptive data-driven reduced-order frameworks capable of operating autonomously together with CFD solvers remains largely unexplored. In realistic online CFD-surrogate scenarios, future reference states are not available a priori, making reliable detection of prediction deterioration a central challenge.

The goal of this work is to develop an adaptive prediction framework that couples an OpenFOAM-based CFD solver with a POD-deep learning (POD-DL) surrogate for long-horizon forecasting under evolving dynamics. The novelty of the proposed framework lies in its online solver-surrogate coupling: instead of relying on precomputed reference trajectories or offline ground-truth data for adaptation, the framework monitors the reliability of the ongoing prediction, recalls the CFD solver only when needed, and updates the surrogate using newly generated high-fidelity snapshots. This makes the method suitable for realistic CFD workflows, where the future flow state is not known a priori and full CFD data are not available over the prediction horizon.

The main contributions are as follows. First, we formulate a closed-loop workflow in which the POD-DL forecasting and OpenFOAM-based data generation are coupled online, so that cumulative autoregressive errors can be corrected by retraining the surrogate with newly generated CFD snapshots. Second, we introduce two control adaptation modes: a prescribed update schedule with user-defined time intervals, and an event-triggered strategy that detects divergence and activates retraining only when needed, improving robustness while limiting the number of expensive solver calls. Third, we assess the framework not only under fixed operating conditions but also under varying inlet velocity, demonstrating its ability to detect regime changes, recall CFD, and recover reliable predictions after model updating. By addressing the lack of ground-truth information during deployment, this work provides a necessary step towards robust and adaptive surrogates for long-horizon CFD prediction.

The remainder of this work is organized as follows. Section 2 presents the proposed methodology, including the adaptive framework, the POD-DL surrogate, the CFD governing equations, and the control logic for online adaptation. Section 3 describes the computational geometry and CFD validation. Section 4 reports results for baseline (non-adaptive prediction) and three adaptive scenarios. Section 5 concludes the paper.

## 2. Methodology

This section presents the proposed adaptive prediction methodology. The following subsections describe the overall adaptive workflow, the governing equations and numerical setup of the CFD solver, the POD-DL model, and the prediction-monitoring metrics.

### *2.1. Adaptive framework*

The proposed framework couples a CFD solver with a POD-DL surrogate to accelerate CFD-based prediction while maintaining long-horizon reliability. Its key feature is online adaptation: the surrogate is used for inexpensive time advancement as long as its prediction remains reliable, whereas high-fidelity CFD is recalled selectively when new information is required. Although OpenFOAM is used here as the CFD solver, the methodology is not tied to a specific CFD code.

Fig. 1 summarizes the closed-loop workflow of the adaptive framework. The process starts by running the CFD solver over an initial training interval to generate snapshots. These snapshots form the initial dataset for training. They are compressed by POD [32], and a sequence predictor is trained in the reduced space to forecast future POD coefficients. Then the simulation enters the adaptive prediction phase. The model uses neural networks (NNs) to perform autoregressive forecasting to advance the reduced state. Depending on the adaptation mode, the prediction is either stopped after a prescribed number of steps or monitored by a divergence-detection module. If the scheduled update time is reached or prediction degradation is detected, the framework switches back to the CFD solver, computes additional high-fidelity snapshots, and retrains or updates the POD-DL predictor before surrogate forecasting is resumed. This alternating process continues until the final time is reached. For clarity and reproducibility, the complete adaptive prediction procedure is summarized in Appendix A.

Two adaptation modes are supported in the present work. The first is a scheduled-update mode, in which retraining is enforced after a prescribed number of time steps. The second is an event-triggered mode, in which retraining is activated automatically when the monitoring criteria persistently exceed a predefined or dynamically estimated threshold. The scheduled mode provides a simple and robust periodic correction, whereas the event-triggered mode reduces unnecessary CFD calls by responding directly to the actual reliability of the ongoing prediction. Together, these two modes provide a balance between prediction robustness and computational cost.

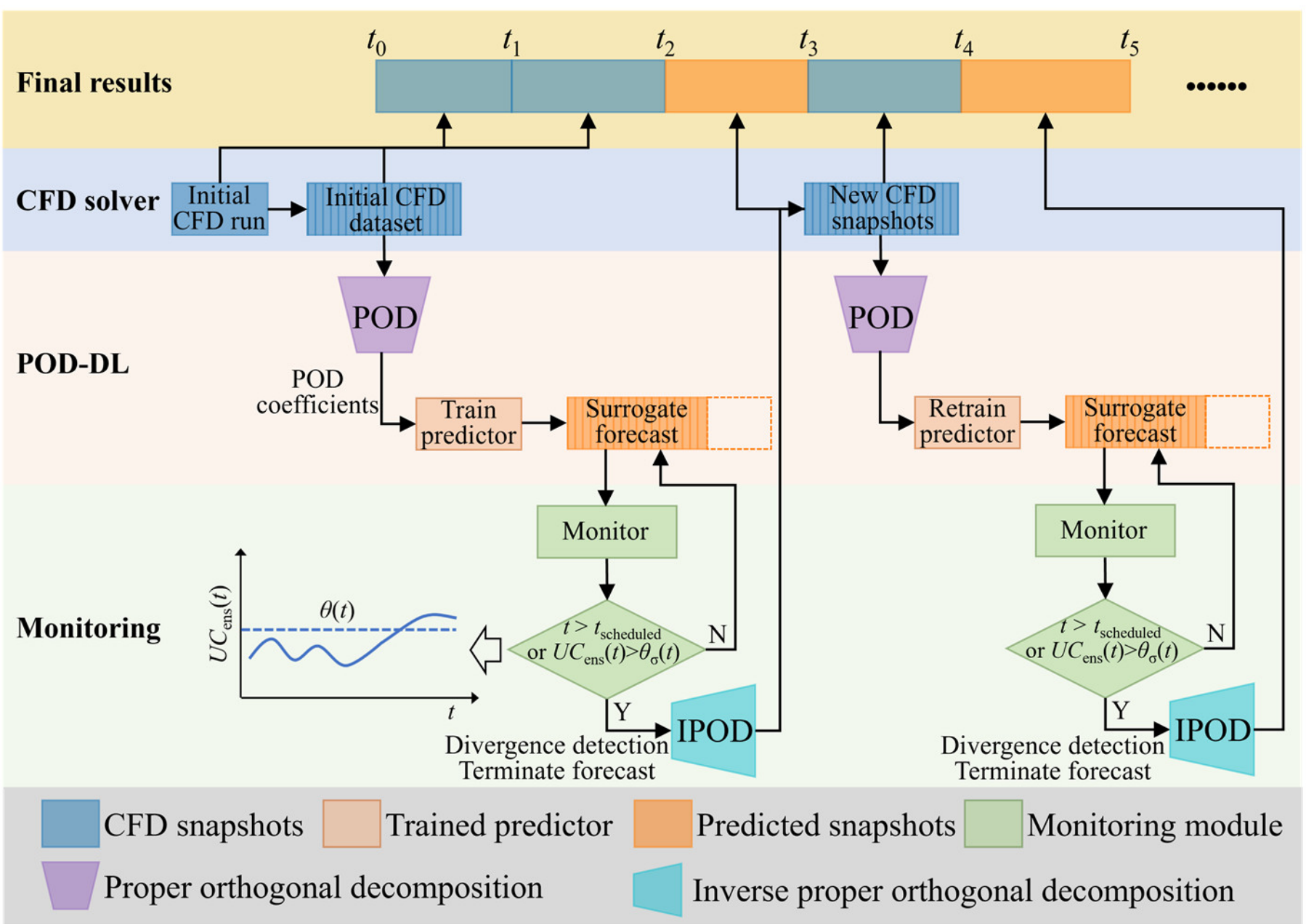

**Fig. 1.** Adaptive framework coupling online CFD simulation and POD-DL prediction.

*2.2. Governing equations of CFD*

High-fidelity data are generated using OpenFOAM [33], an open-source finite-volume CFD toolbox. In this study the flow is modelled as an incompressible Newtonian fluid governed by the Navier-Stokes equations,

$$\nabla \cdot \mathbf{U} = 0, \tag{1}$$

$$\frac{\partial \mathbf{U}}{\partial t} + \mathbf{U} \cdot \nabla \mathbf{U} = -\nabla p + \nu \Delta \mathbf{U}, \tag{2}$$

where $\mathbf{U}$ is the velocity, $p$ is the kinematic pressure, and $\nu$ is the kinematic viscosity. The Reynolds number is defined as

$$\mathrm{Re} = U_{\mathrm{in}} D / \nu, \tag{3}$$

where $U_{\mathrm{in}}$ is the reference inlet velocity and $D$ is the cylinder diameter.

In this work, the flow is solved without a turbulence model, and the kinematic viscosity is prescribed to achieve the target Reynolds number. For $U_{\mathrm{in}} = 1$ m/s, $\nu = 5\times10^{-3}$ m$^2$/s and $3.33\times10^{-3}$ m$^2$/s are used for Re = 200 and 300, respectively. The boundary conditions are specified as follows. At the inlet, the velocity is fixed to $U_{in}$, while the pressure uses a zero-gradient condition. On the cylinder surface, a no-slip condition is imposed for velocity, and pressure is treated with a zero-gradient condition. At the outlet, the velocity is assigned a mixed outflow/backflow treatment: a zero-gradient condition is applied for outward flow, whereas a prescribed reference velocity is imposed only if local backflow occurs. The pressure is fixed to a reference value at the outlet. The spanwise boundaries are periodic.

The governing equations are discretized in OpenFOAM [33] using the finite-volume method. Second-order schemes are employed for the spatial discretization of the gradient and divergence terms, and time integration is performed using the implicit second-order Crank-Nicolson scheme. The resulting linear systems are solved using a GAMG solver for pressure and a PBiCGStab solver with DILU preconditioning for velocity. Pressure-velocity coupling is handled using the PIMPLE algorithm [33,34].

For validation and quantitative assessment, standard integral quantities are used, including the time-averaged drag and base-pressure coefficients. The drag and base-pressure coefficients are respectively defined as

$$C_{\mathrm{d}}(t) = \frac{F_{\mathrm{d}}}{\frac{1}{2}\rho U_{\mathrm{in}}^2 A_{\mathrm{ref}}}, \tag{4}$$

$$C_{\mathrm{pb}}(t) = \frac{p_{\mathrm{b}} - p_{\infty}}{\frac{1}{2}U_{\mathrm{in}}^2}, \tag{5}$$

where $F_{\mathrm{d}}$ is the drag force, $A_{\mathrm{ref}} = D \cdot L$ is the reference area, $L$ is the spanwise length, $p_{\mathrm{b}}$ is the base pressure, and $p_{\infty}$ is the reference pressure. Therefore, the time-averaged $C_{\mathrm{d}}$ and $C_{\mathrm{pb}}$ are respectively calculated as

$$\bar{C}_{\mathrm{d}} = \frac{1}{N_{\mathrm{snap}}} \sum_{n=1}^{N_{\mathrm{snap}}} C_{\mathrm{d}}\left(t_{\mathrm{n}}\right), \tag{6}$$

$$\bar{C}_{\mathrm{pb}} = \frac{1}{N_{\mathrm{snap}}} \sum_{n=1}^{N_{\mathrm{snap}}} C_{\mathrm{pb}}\left(t_{\mathrm{n}}\right). \tag{7}$$

The vortex-shedding frequency is expressed by the Strouhal number,

$$\mathrm{St}=\frac{f_{\mathrm{s}}D}{U_{\mathrm{in}}}, \tag{8}$$

where $f_s$ is the shedding frequency.

To better reveal the temporal trends of these coefficients, the original signals are smoothed using a moving average. Specifically, the smoothed $C_d$ at snapshot $i$ is computed as

$$\tilde{C}_{\mathrm{d,i}}=\frac{1}{2m+1}\sum_{k=-m}^{m}C_{\mathrm{d,i+k}} \tag{9}$$

where $C_{d,i+k}$ is the original drag coefficient at snapshot $i+k$, and $2m+1$ is the smoothing-window length. At the beginning and end of the curve, edge padding is applied so that the smoothed and original sequences have the same length. The same operation is applied to lift coefficient $C_l$.

*2.3. POD-DL model*

The CFD solution is sampled at discrete times to form a snapshot dataset. For a structured grid, the unsteady flow field can be written as a fifth-order tensor, represented by

$$\mathbf{X}\in\mathbb{R}^{C\times N_{\mathrm{x}}\times N_{\mathrm{y}}\times N_{\mathrm{z}}\times K}, \tag{10}$$

where $C$ denotes the number of variable components, $N_x$, $N_y$, $N_z$ represent the grid sizes in the streamwise, cross-stream and spanwise directions, and $K$ is the number of snapshots. More generally, the spatial coordinates can be arranged into a single dimension when the grid cannot be naturally organized in separate directions, as in the case of unstructured meshes. In the present work, different variables are predicted in parallel and processed separately. Therefore, the tensor **X** is reshaped by stacking spatial degrees of freedom, given as

$$\mathbf{A}\in\mathbb{R}^{C\times J\times K}, J=N_{\mathrm{x}}\times N_{\mathrm{y}}\times N_{\mathrm{z}}. \tag{11}$$

Before POD decomposition, the snapshot matrix is mean-centered by subtracting the temporal mean at each spatial degree of freedom, expressed as

$$\mathbf{A}'=\mathbf{A}-\overline{\mathbf{A}},\ \overline{\mathbf{A}}=\frac{1}{K}\sum_{k=1}^{K}\mathbf{A}_{:,\mathrm{k}}. \tag{12}$$

where $\overline{\mathbf{A}}$ is the temporal mean field and $\mathbf{A}'$ is the fluctuation matrix. This removes the stationary component of the flow and allows the POD basis to focus on coherent unsteady fluctuations. As one of the methods to perform POD, singular value decomposition (SVD) is used to conduct dimensionality reduction [32] for every variable component, given as

$$\mathbf{A}' = \mathbf{\Phi\Sigma V}^{\mathrm{T}}, \tag{13}$$

where $\mathbf{\Phi}$ contains spatial POD modes of a component, $\mathbf{V}$ contains temporal coefficients, and $\mathbf{\Sigma} = \mathbf{diag}(\sigma_1, \sigma_2, ...)$ stores singular values. A truncated basis of rank $r$ is selected using a cumulative energy criterion based on singular values. The reduced representation is then reconstructed using the leading $r$ modes,

$$\mathbf{A}' \approx \mathbf{\Phi}_{\mathrm{r}}\mathbf{\Sigma}_{\mathrm{r}}\mathbf{V}_{\mathrm{r}}^{\mathrm{T}}, \tag{14}$$

where $\mathbf{V}_{\mathrm{r}}^{\mathrm{T}}$ contains the truncated temporal coefficients. For the temporal coefficient matrix $\mathbf{V}_{\mathrm{r}}^{\mathrm{T}}$, each retained modal coefficient is standardized as

$$\tilde{c}_{\mathrm{i}}(t_{\mathrm{k}}) = \frac{v_{\mathrm{i}}(t_{\mathrm{k}}) - \mu_{\mathrm{i}}}{s_{\mathrm{i}} + \varepsilon}, i = 1, 2, ..., r \tag{15}$$

where $v_{\mathrm{i}}(t_{\mathrm{k}})$ is the temporal coefficient of the $i$-th retained POD mode at snapshot $k$, $\mu_{\mathrm{i}}$ and $s_{\mathrm{i}}$ are the temporal mean and standard deviation of the $i$-th POD coefficient over the training snapshots, and $\varepsilon = 10^{-7}$ is added for numerical stability. This preprocessing separates the steady mean field from the unsteady dynamics, balances the magnitudes of different modal coefficients, and improves numerical conditioning for the neural-network predictor. The same procedure is applied separately to each variable component.

Let $\mathbf{c}_{\mathrm{t}} \in \mathbb{R}^{\mathrm{r}}$ denote the vector of retained POD coefficients at time $t$. The temporal predictor is trained in a supervised manner using rolling windows of the coefficient sequence. Specifically, a sequence of $q$ consecutive coefficient vectors is used as input, and the following $p$ coefficient vectors are used as the prediction target,

$$\left[\hat{\mathbf{c}}_{\mathrm{t+1}}, \hat{\mathbf{c}}_{\mathrm{t+2}}, ..., \hat{\mathbf{c}}_{\mathrm{t+p}}\right] = f_{\theta}\left(\mathbf{c}_{\mathrm{t-q+1}}, \mathbf{c}_{\mathrm{t-q+2}}, ..., \mathbf{c}_{\mathrm{t}}\right), \tag{16}$$

where $q$ is the input sequence length, $p$ is the output sequence length, and $f_{\theta}$ is the neural-network predictor parameterized by $\theta$. In the current implementation, $f_{\theta}$ is a stacked long short-term memory (LSTM) network followed by fully connected layers and a linear output layer. The model is trained by minimizing the mean-squared error between the predicted and target POD coefficients, using mini-batch optimization with the Adam optimizer [14].

During prediction, the trained predictor is executed autoregressively. Starting from the most recent $q$ coefficient vectors, the model predicts the next block of $p$ coefficient vectors. The rolling input window is then updated by removing the oldest entries and appending the newly predicted coefficients. This procedure is repeated to obtain long-horizon forecasts. The predicted coefficient matrix $\hat{\mathbf{C}}$ is transformed back to the physical space through the inverse POD reconstruction, after which the temporal mean field is added to obtain the final predicted variable component,

$$\hat{\mathbf{A}} \simeq \overline{\mathbf{A}} + \mathbf{\Phi}_{\mathrm{r}} \mathbf{\Sigma}_{\mathrm{r}} \hat{\mathbf{C}}, \tag{17}$$

If an external control schedule is available, such as a varying inlet velocity, the normalized control value is concatenated with the POD coefficient vector at each input time step, enabling conditional forecasting under varying operating conditions.

If the ground truth data are available, the prediction error is evaluated using the root mean square error (RMSE),

$$\mathrm{RMSE} = \left\| \mathbf{A} - \hat{\mathbf{A}} \right\|_2, \tag{18}$$

where $\mathbf{A}$ and $\hat{\mathbf{A}}$ denote the reference and predicted fields, respectively. RMSE is adopted because it directly measures the overall magnitude of the field-wise prediction error and gives stronger weight to large local deviations, making it suitable for evaluating accumulated errors in long-horizon flow prediction.

In addition to RMSE, turbulent kinetic energy (TKE) and temporal correlation are used to assess whether the surrogate preserves the main unsteady wake dynamics. The TKE is computed from the three velocity components as

$$\mathrm{TKE} = \frac{1}{2}\left( \left( U_{\mathrm{x}} - \overline{U_{\mathrm{x}}} \right)^2 + \left( U_{\mathrm{y}} - \overline{U_{\mathrm{y}}} \right)^2 + \left( U_{\mathrm{z}} - \overline{U_{\mathrm{z}}} \right)^2 \right), \tag{19}$$

where $U_{\mathrm{x}}$, $U_{\mathrm{y}}$, and $U_{\mathrm{z}}$ are the instantaneous velocity components in streamwise, cross-stream and spanwise directions, respectively, and the overbar denotes temporal averaging. This quantity is used to evaluate whether the prediction retains the energetic distribution of the unsteady wake.

The temporal correlation of the streamwise velocity is defined as

$$R_{\mathrm{uu}}(t)=\frac{\left\langle U_{\mathrm{x}}'(t)U_{\mathrm{x}}'(t+\tau)\right\rangle_{\mathrm{t}}}{\left\langle U_{\mathrm{x}}'^{2}(t)\right\rangle_{\mathrm{t}}},U_{\mathrm{x}}'(t)=U_{\mathrm{x}}(t)-\left\langle U_{\mathrm{x}}(t)\right\rangle_{\mathrm{t}}, \tag{20}$$

where $U_{\mathrm{x}}(t)$ is the streamwise velocity signal at a selected probe location, $U_{\mathrm{x}}'(t)$ is its fluctuation, $\tau$ is the time lag, and $\langle\cdot\rangle_{\mathrm{t}}$ denotes temporal averaging.

*2.4. Prediction-monitoring and evaluation metrics*

Adaptive prediction requires a lightweight but informative mechanism to determine whether the surrogate remains within a trustworthy regime. In the proposed framework, monitoring is performed primarily in the reduced space, where deviations from the training manifold and abnormal redistribution of modal energy can be evaluated at negligible cost compared with CFD. Ensemble-based uncertainty indicators are used to quantify prediction uncertainty and detect unreliable forecasts. Multiple surrogate models trained with different random seeds are advanced simultaneously, and their spread is used to estimate ensemble uncertainty. This uncertainty indicator is then used as an online trigger for adaptation.

Let $\left\{\hat{\mathbf{c}}^{(m)}(t)\right\}_{m=1}^{M}$ denote the ensemble forecasts of the retained POD coefficient vector at time $t$, where $M$ is the number of predictors. The ensemble mean $\overline{\mathbf{c}}(t)$ and per-mode standard deviation $s_{\mathrm{i}}(t)$ are defined as

$$\overline{\mathbf{c}}(t)=\frac{1}{M}\sum_{m=1}^{M}\hat{\mathbf{c}}^{(m)}(t), \tag{21}$$

$$s_{\mathrm{i}}(t)=\sqrt{\frac{1}{M-1}\sum_{m=1}^{M}\left(\hat{c}_{\mathrm{i}}^{(m)}(t)-\overline{c}_{\mathrm{i}}(t)\right)^{2}},i=1,2,\ldots,r. \tag{22}$$

A scalar ensemble-uncertainty indicator is computed as an energy-weighted measure of the spread among ensemble predictions in the POD coefficient space. The per-mode ensemble standard deviation is weighted by the corresponding retained singular value, so that uncertainty in energetically dominant modes contributes more strongly to the final indicator. And it is computed as

$$UC_{\mathrm{ens}}(t)=\sqrt{\frac{\sum_{i=1}^{r}\sigma_{\mathrm{i}}^{2}\left(s_{\mathrm{i}}^{\mathrm{ens}}(t)\right)^{2}}{\sum_{i=1}^{r}\sigma_{\mathrm{i}}^{2}+\varepsilon}}, \tag{23}$$

where $\sigma_{\mathrm{i}}$ denotes the singular value of the $i$-th retained mode, $s_{\mathrm{i}}^{\mathrm{ens}}(t)$ is the ensembles standard deviation of the $i$-th POD coefficient at time $t$, and $\varepsilon = 10^{-7}$ is a small constant used for numerical stability. This energy-weighted definition gives importance to uncertainty in energetically dominant modes.

In the present implementation, the threshold for $UC_{\mathrm{ens}}$ can be either prescribed externally or estimated online. If an external threshold is provided, it is directly used as

$$\theta_{\sigma}(t)=\theta_{\sigma,\mathrm{pres}}. \tag{24}$$

Otherwise, an adaptive threshold is computed from a sliding buffer of recent uncertainty values. Since the prediction is performed at discrete steps, let $t_{\mathrm{j}}$ denote the $j$-th prediction step and $UC_{\mathrm{ens}}(t_{\mathrm{j}})$ the corresponding ensemble uncertainty. The sliding buffer is defined as

$$B_{\sigma}(t_{\mathrm{j}})=\left\{UC_{\mathrm{ens}}(t_{\mathrm{l}})\,|\,l=\max(1,j-N_{\sigma}+1),\dots,j)\right\} \tag{25}$$

where $N_{\sigma}$ is the window length. No uncertainty-based trigger is allowed until the buffer contains at least $N_{\mathrm{min}} = 20$ samples. Once this condition is satisfied, the threshold is estimated using the median-MAD rule,

$$m_{\sigma}(t_{\mathrm{j}})=\mathrm{median}\left(B_{\sigma}(t_{\mathrm{j}})\right), \tag{26}$$

$$\mathrm{MAD}_{\sigma}(t_{\mathrm{j}})=\mathrm{median}\left(\left|UC_{\mathrm{ens}}(t_{\mathrm{l}})-m_{\sigma}(t_{\mathrm{j}})\right|\right), \tag{27}$$

$$\theta_{\sigma}(t_{\mathrm{j}})=m_{\sigma}(t_{\mathrm{j}})+1.4826\cdot k_{\sigma}\cdot\mathrm{MAD}_{\sigma}(t_{\mathrm{j}}), \tag{28}$$

where the factor 1.4826 makes the MAD estimate consistent with the standard deviation for a Gaussian distribution. The coefficient $k_{\sigma}$ is a threshold multiplier that controls the sensitivity of the divergence detector: smaller values lead to earlier triggering, whereas larger values make the criterion more conservative. In this study, $k_{\sigma} = 3.0$ is adopted, corresponding to a three-sigma-type rule, so that CFD is recalled only when the ensemble uncertainty exceeds the local baseline by a sufficiently large margin. The prediction at step $t_{\mathrm{j}}$ is marked as unreliable when

$$UC_{\text{ens}}\left(t_{\text{j}}\right) > \theta_{\sigma}\left(t_{\text{j}}\right). \tag{29}$$

The prediction is truncated only when the abnormality condition persists for $N_{\text{consec}}$ consecutive steps. In the present study, $N_{\text{consec}} = 2$. Once triggered, the retained part of the prediction is taken as the reliable segment, and the CFD solver is recalled from the corresponding state to generate new high-fidelity snapshots.

## 3. Computational geometry and model validation

The proposed adaptive framework is evaluated based on the canonical problem of flow around a circular cylinder. High-fidelity simulation was performed with OpenFOAM [33] using pimpleFoam for Re ranging from 200 to 400. The accuracy of such simulations remains sensitive to mesh resolution, time step, spanwise degrees of freedom, and blockage effects induced by the far-field boundaries. Therefore, the numerical setup must be validated to ensure that the simulated results are physically reliable before conducting prediction.

A three-dimensional configuration is considered, with a diameter of circular cylinder of $D$, a domain height of $20D$ and a spanwise length of $3D$. The inlet is located $10D$ upstream of the cylinder while the downstream length is $20D$, which is suitable for this case [35]. The computational geometry and mesh are shown in Fig. 2, where the mesh on the vertical plane is displayed for clarity. The total number of cells used in this study is 741 thousand, and the mesh is refined near the cylinder and in the wake region. Different Reynolds numbers are obtained by adjusting the kinematic viscosity or inlet velocity while keeping the geometric configuration unchanged. This strategy provides a consistent benchmark set for assessing the surrogate performance under different wake regimes without introducing additional geometric variability. Additionally, the location of probes is plotted in Fig. 2 (b), which is used to evaluate the prediction in the following Section.

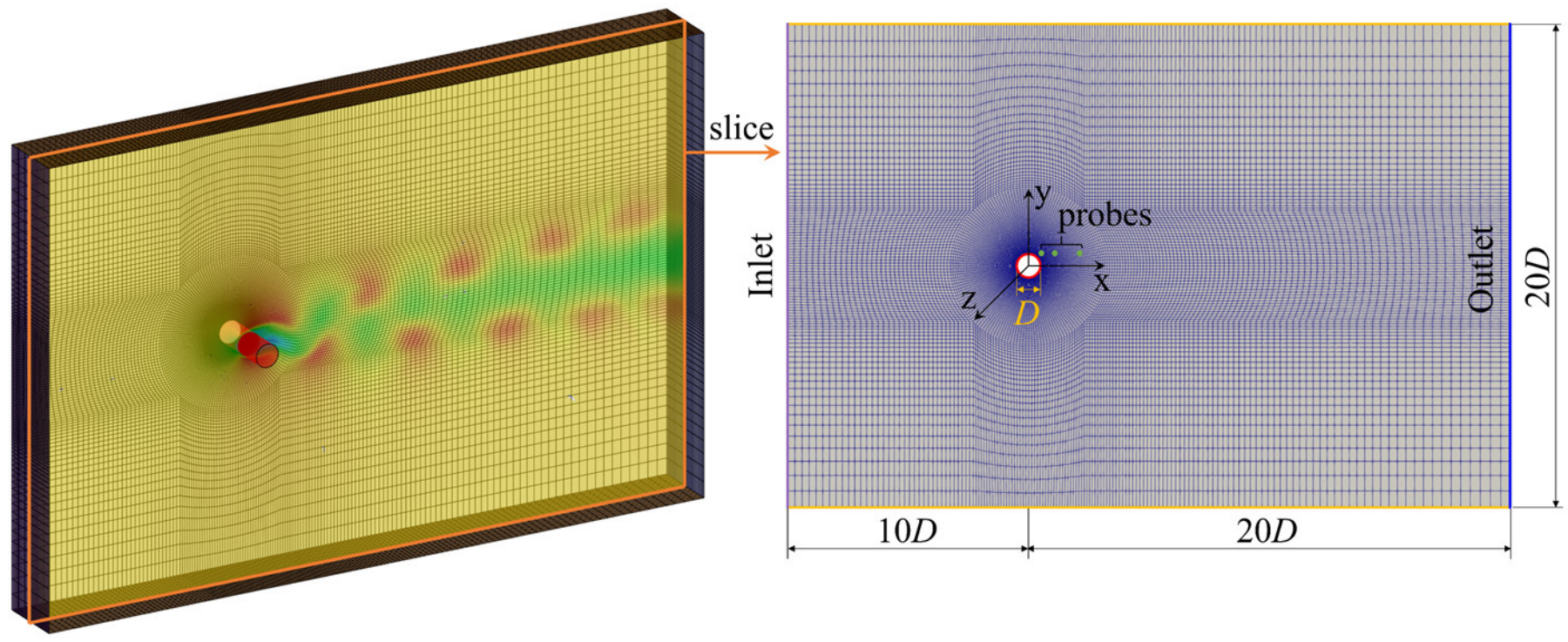

**Fig. 2.** Computational geometry and mesh for flow past a circular cylinder.

To validate the numerical setup, the mean drag coefficient $\overline{C}_{\mathrm{d}}$, the mean base-pressure coefficient $\overline{C}_{\mathrm{pb}}$ and the Strouhal number *St* are compared against experimental data [36–38] for Re = 200–400 in Fig. 3. The comparison shows that $\overline{C}_{\mathrm{d}}$ decreases gradually as Re increases, while $\overline{C}_{\mathrm{pb}}$ exhibits an increasing trend over the same range. The shedding frequency, expressed through *St*, remains close to 0.2 for Re ≤ 250 and increases for Re ≥ 250, which is consistent with the well-known behavior of cylinder wakes in this Reynolds number range. Overall, the numerical predictions follow the experimental trends closely, demonstrating that the present OpenFOAM setup captures the dominant flow physics with satisfactory fidelity. This agreement provides a suitable foundation for the subsequent development and evaluation of the POD-DL model.

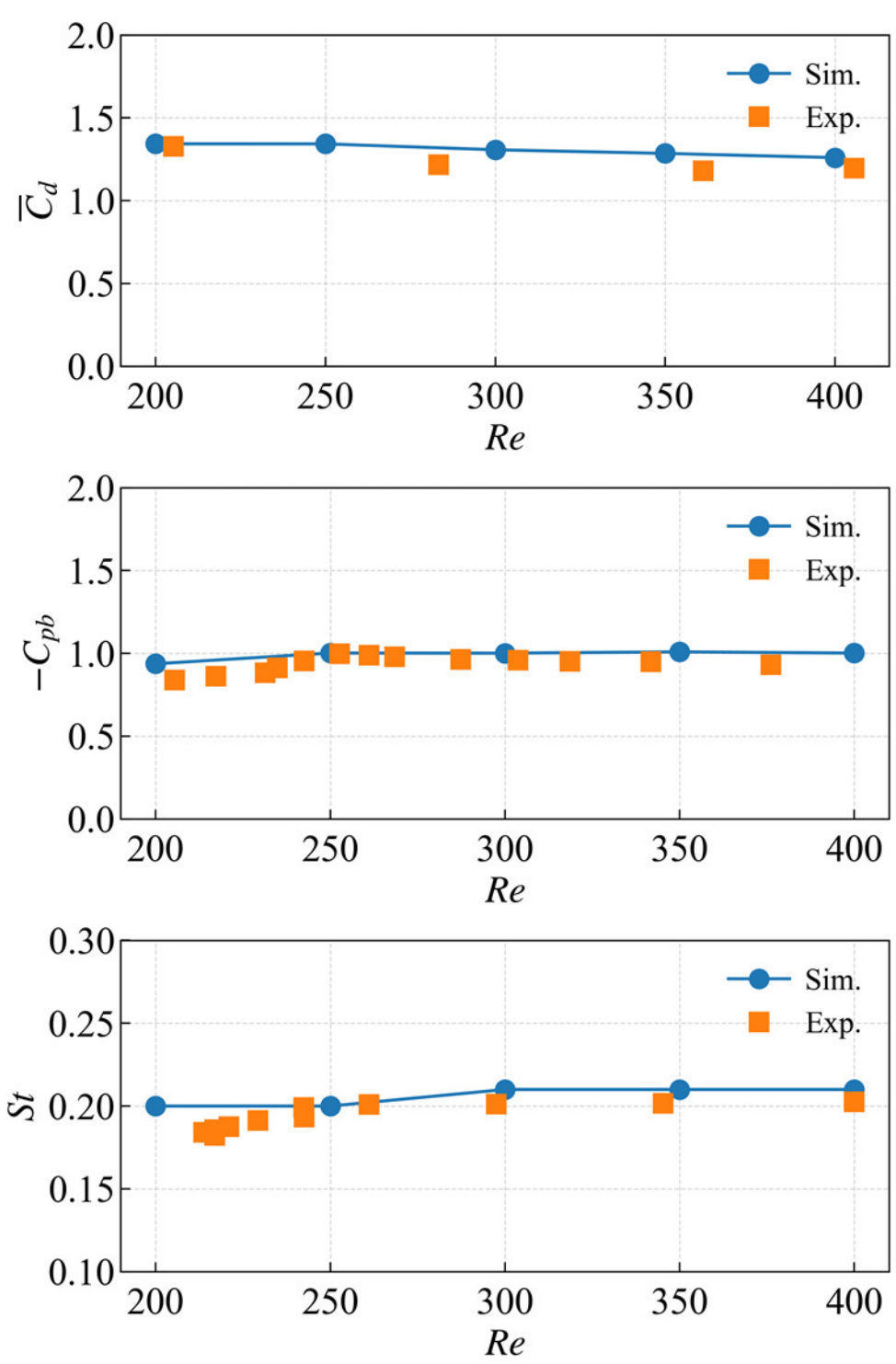

**Fig. 3.** Comparison of mean drag coefficient $\overline{C}_{\mathrm{d}}$, mean base-pressure coefficient $\overline{C}_{\mathrm{pb}}$, and Strouhal number *St* between the present CFD simulations and experiments [36–38].

## 4. Results and discussion

This section presents and discusses the prediction results. The baseline results without adaptation are first examined to illustrate the error accumulation of the original POD-DL surrogate. Three adaptive strategies are then analyzed, including scheduled updates, event-triggered adaptation, and adaptive prediction under varying inlet velocity.

### *4.1. Baseline prediction without adaptation*

We first examine the baseline, non-adaptive prediction behavior for two representative cases. The purpose of this section is to assess how well the POD-DL surrogate reproduces the wake dynamics when trained on an initial CFD dataset, and to identify the progressive degradation mechanism that motivates the adaptive framework.

The first case corresponds to Re = 200, representing an organized unsteady wake regime. The wake dynamics are analyzed based on a representative time window 300-305 s, which covers one vortex shedding period from the cylinder wake. The streamwise and cross-stream velocity components in the slice plane ($z = 0$) are shown in Fig. 4. A separation region forms immediately behind the cylinder, from which vortices are shed alternately into the downstream wake. In the $U_x$ field, the separated wake is identified by the low-speed region along the wake centreline, while elongated high- and low-speed structures appear alternately on both sides of the wake. The periodic organization is more evident in the $U_y$ field, where alternating positive and negative lobes persist from the near wake to the intermediate wake, forming a clear antisymmetric pattern associated with the Kármán vortex street. These structures move downstream with nearly uniform spacing and without strong deformation, indicating a nearly periodic shedding process.

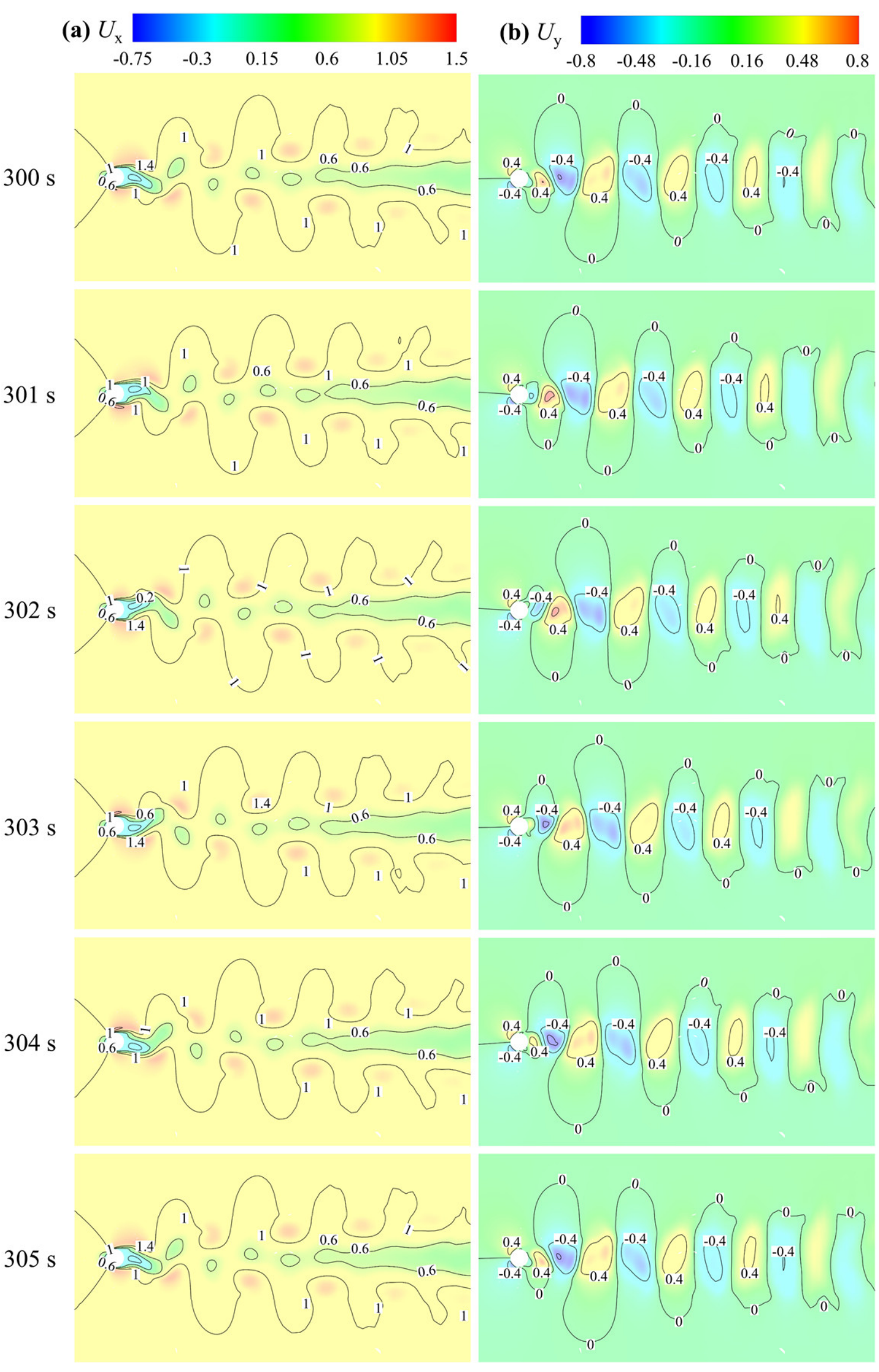

**Fig. 4.** Instantaneous (a) streamwise velocity ($U_x$) and (b) cross-stream velocity ($U_y$) fields for the flow past a circular cylinder at Re = 200 over a representative period.

The initial transient of the CFD simulation is discarded before constructing the dataset. Specifically, the first 100 s are excluded, and $t$ = 100 s is redefined as snapshot 0. This avoids training the POD-DL model on non-representative dynamics and ensures that the surrogate focuses on the established vortex-shedding regime. With a sampling step of 0.5 s, the first 200 snapshots after this re-indexing are used for training, and the trained model is then employed to predict the subsequent 600 snapshots.

The modal characteristics of the training dataset are examined through the singular-value decay shown in Fig. 5. The cumulative energy ratio, defined in our previous work [18], increases rapidly with the number of retained modes for all variables, indicating that 25 POD modes capture more than 80% of the resolved energy. The normalized singular values in Fig. 5 (b) also show a relatively rapid decay, suggesting that the dominant wake dynamics are concentrated in the leading modes. This behavior is favorable for reduced-order prediction, because the essential unsteady flow features are governed by a low-dimensional set of coherent structures.

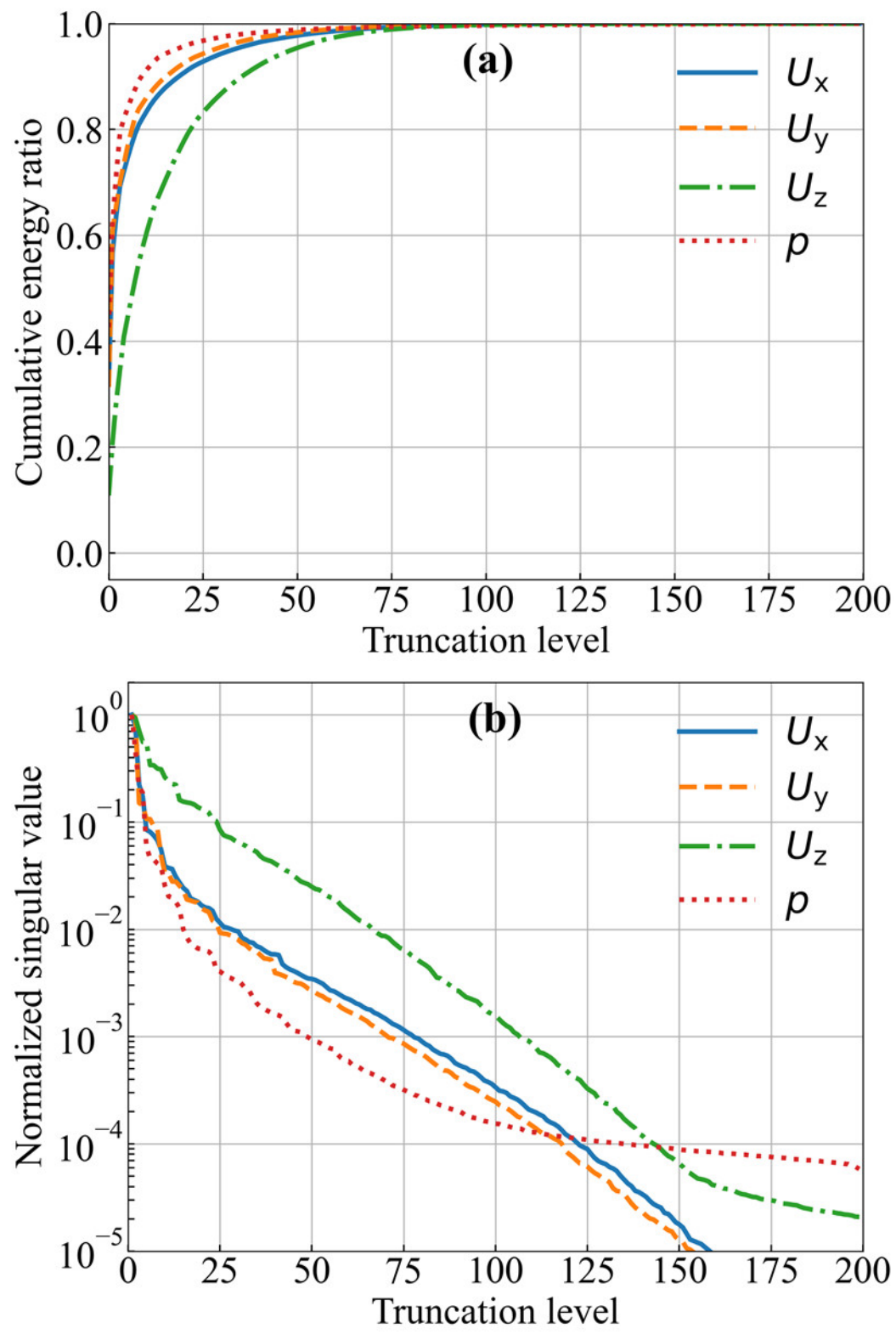

**Fig. 5.** (a) Cumulative energy ratio and (b) normalized singular values of POD modes for the flow

past a circular cylinder at Re = 200.

A representative probe location (0.5$D$, 0.5$D$, 0) as shown in Fig. 2 is selected to compare the variable evolution between the CFD ground truth and the predicted results, which is shown in Fig. 6. For $U_x$, $U_y$ and $p$, the predicted variations follow the oscillatory trends of the CFD data well in both amplitude and phase over a substantial portion of the forecasting window. For $U_z$, the signal rapidly decays towards a near-zero level, and the prediction follows this behavior closely. Overall, these results indicate that the POD-DL model successfully captures the main oscillatory dynamics and produces physically meaningful long-horizon forecasts in this flow regime.

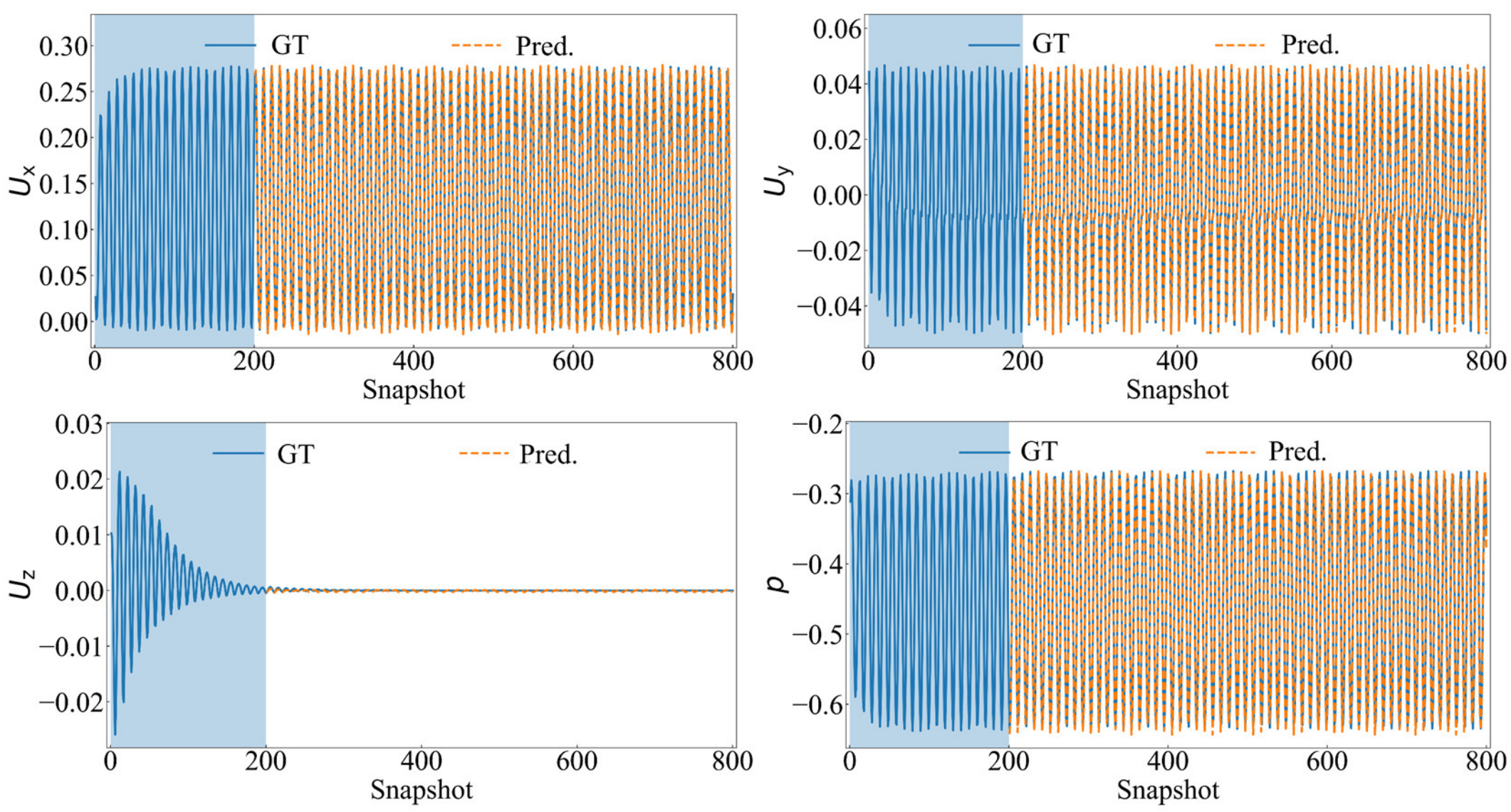

**Fig. 6.** Comparison of $U_x$, $U_y$, $U_z$ and $p$ between prediction and CFD ground truth at probe location (0.5$D$, 0.5$D$, 0) for the flow past a circular cylinder at Re=200.

Nevertheless, the root mean square error (RMSE), shown in Fig. 7, increases gradually as the prediction advances. This trend demonstrates that the prediction accuracy degrades progressively as the forecast advances. Even when the short-term agreement is satisfactory, autoregressive forecasting inevitably accumulates small state errors, which are then propagated and amplified over time. In other words, the baseline surrogate remains accurate over a finite horizon, but its reliability deteriorates progressively as the forecast is extended. This behavior motivates adaptive prediction: even in this relatively regular wake regime, a fixed offline-trained model cannot remain uniformly reliable over long forecast horizons.

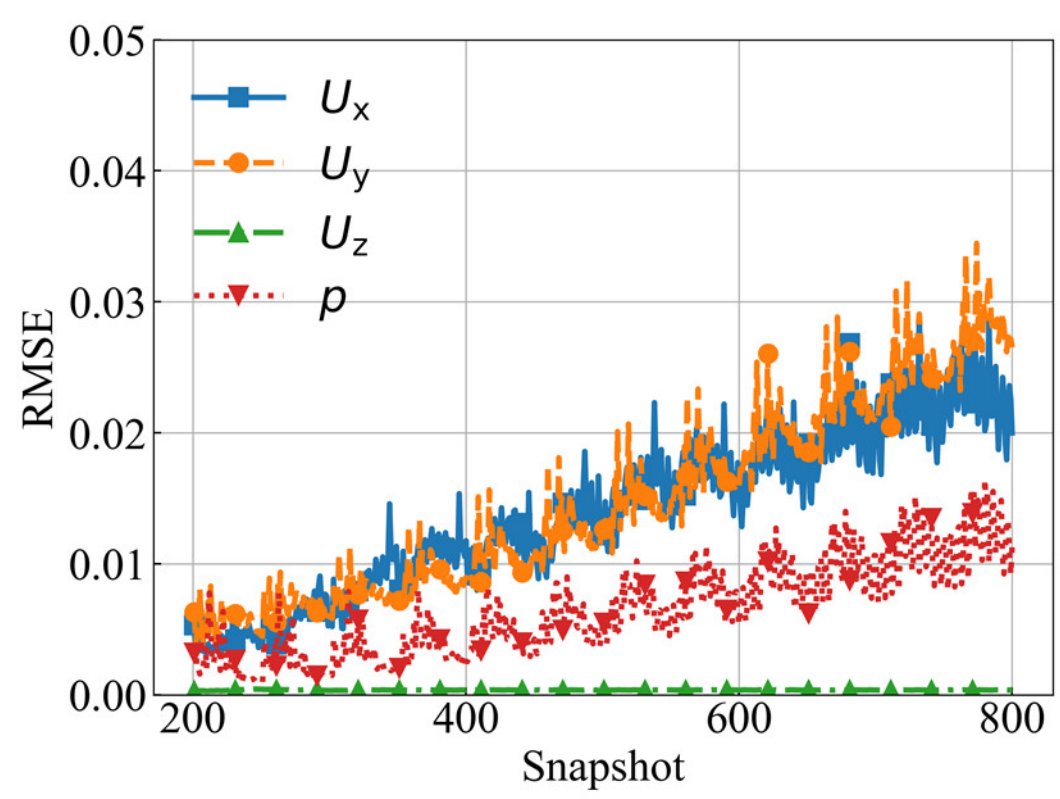

**Fig. 7.** Root mean square error (RMSE) between prediction and CFD ground truth for the flow past a circular cylinder at Re=200.

A second baseline case is considered at Re = 300. The instantaneous velocity fields in the vertical plane ($z$ = 0) are shown in Fig. 8 over the same time window. Compared with the Re = 200 case, the wake still exhibits alternating vortex shedding, but the separated shear layers and downstream vortical structures become more complex. In the $U_x$ field, the wake shows stronger spatial modulation and more pronounced local deformation. In the $U_y$ field, alternating positive and negative lobes are still visible, but their shape and spacing become less uniform. This indicates that the wake dynamics at Re = 300 are more difficult to represent and predict.

The cumulative energy ratio and mode decay of singular values are shown in Fig. 9. When 25 POD modes are retained, the cumulative energy ratios of $U_x$, $U_y$, $U_z$ and $p$ are 56%, 60%, 40% and 70%, respectively. Compared with the results of Re = 200, these values are lower, indicating that the energetic content at Re = 300 is distributed over a broader range of modes, especially for the spanwise velocity $U_z$. Although the retained energy is lower than in the Re = 200 case, the leading 25 modes still capture a substantial fraction of the energetic content and preserve the dominant coherent wake structures, as demonstrated in Ref. [16]. Therefore, the POD representation remains suitable for the reduced-order prediction. However, compared with the lower-Re case, the energy is distributed over a larger number of modes, so the subsequent prediction step becomes more challenging.

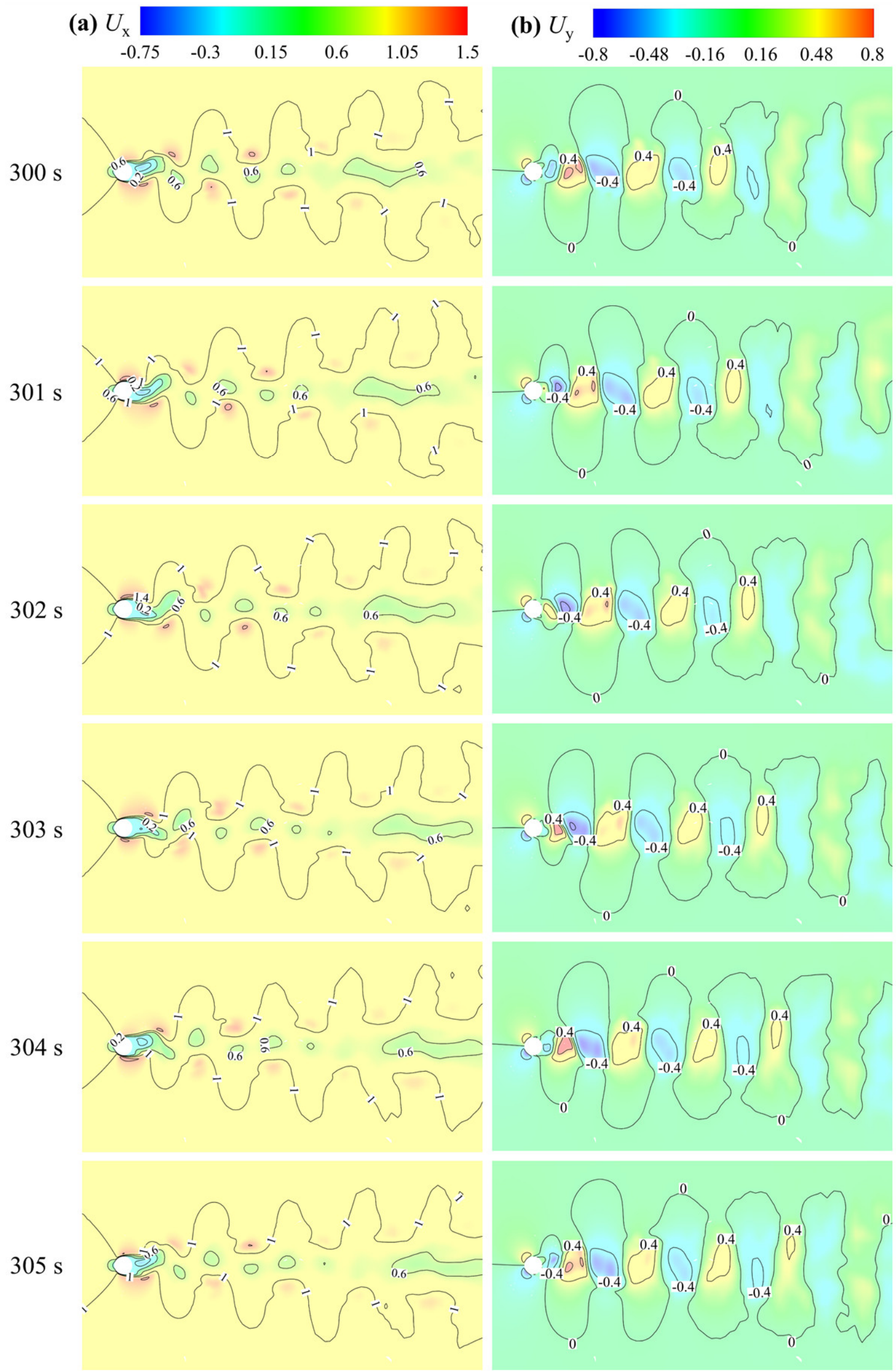

**Fig. 8.** Counterpart of Fig. 4 for Re = 300, showing the instantaneous $U_x$ and $U_y$ fields over the same representative shedding period.

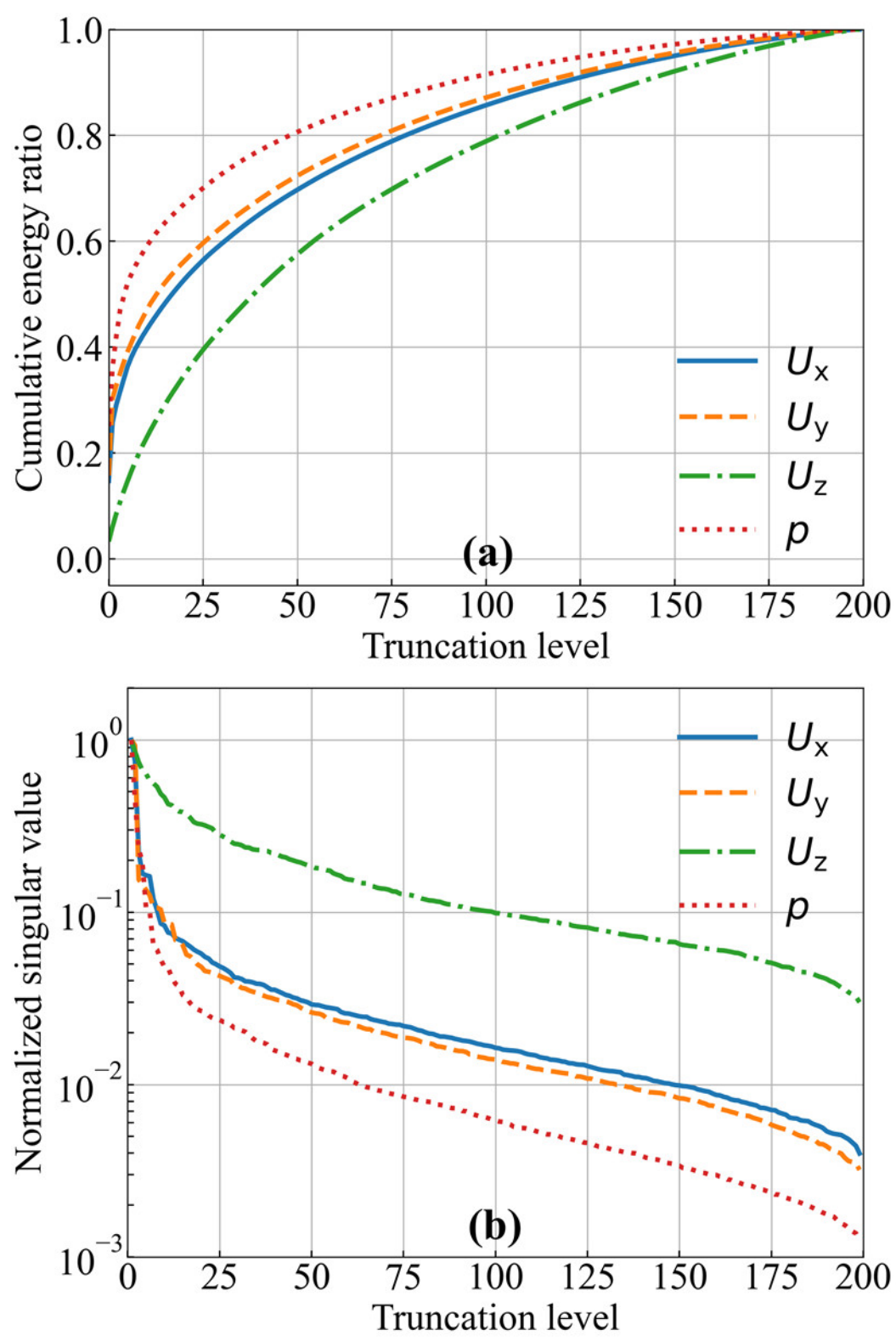

**Fig. 9.** Counterpart of Fig. 5 for Re = 300, showing the cumulative energy ratio and normalized singular values of the POD modes.

The comparison between prediction and ground truth at probe (0.5$D$, 0.5$D$, 0) is shown in Fig. 10. The prediction captures the dominant oscillatory behavior and reproduces the overall temporal patterns of the wake, which confirms that the retained POD modes preserve the main coherent structures of the flow. However, compared with the laminar case, the mismatch relative to CFD becomes more noticeable as the prediction advances. The phase position agreement is less stable, and the amplitudes of the oscillations are reproduced with lower consistency over long forecast horizons. This behavior is consistent with the broader modal spectrum and the increased dynamical complexity at Re = 300.

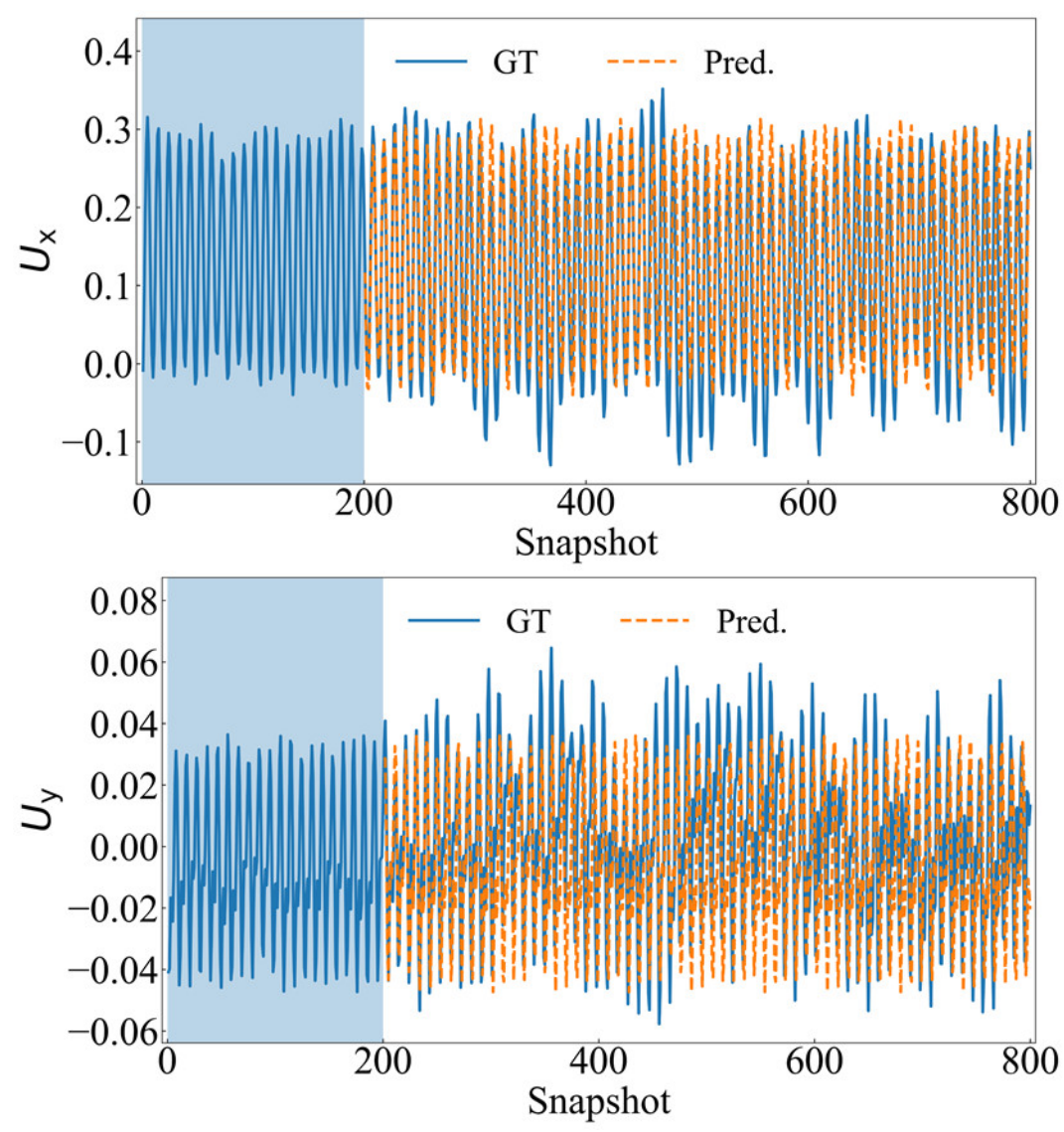

**Fig. 10.** Comparison of streamwise and cross-stream velocity between prediction and CFD ground truth at the probe location (0.5*D*, 0.5*D*, 0) for the flow past a circular cylinder at Re = 300.

The RMSE between the prediction and CFD ground truth at Re = 300 is shown in Fig. 11, which further illustrates the increased difficulty of long-horizon prediction in this regime. First, the error level is clearly higher than for Re = 200. Second, the overall trend remains near-monotonic growth with increasing snapshot index, demonstrating that error accumulation increases as the prediction advances. These results show that the baseline POD-DL model, while effective for short- to intermediate-horizon forecasting, is not sufficiently robust to maintain stable accuracy over long horizons when the dynamics become more complex. This observation directly motivates the introduction of the online adaptive framework.

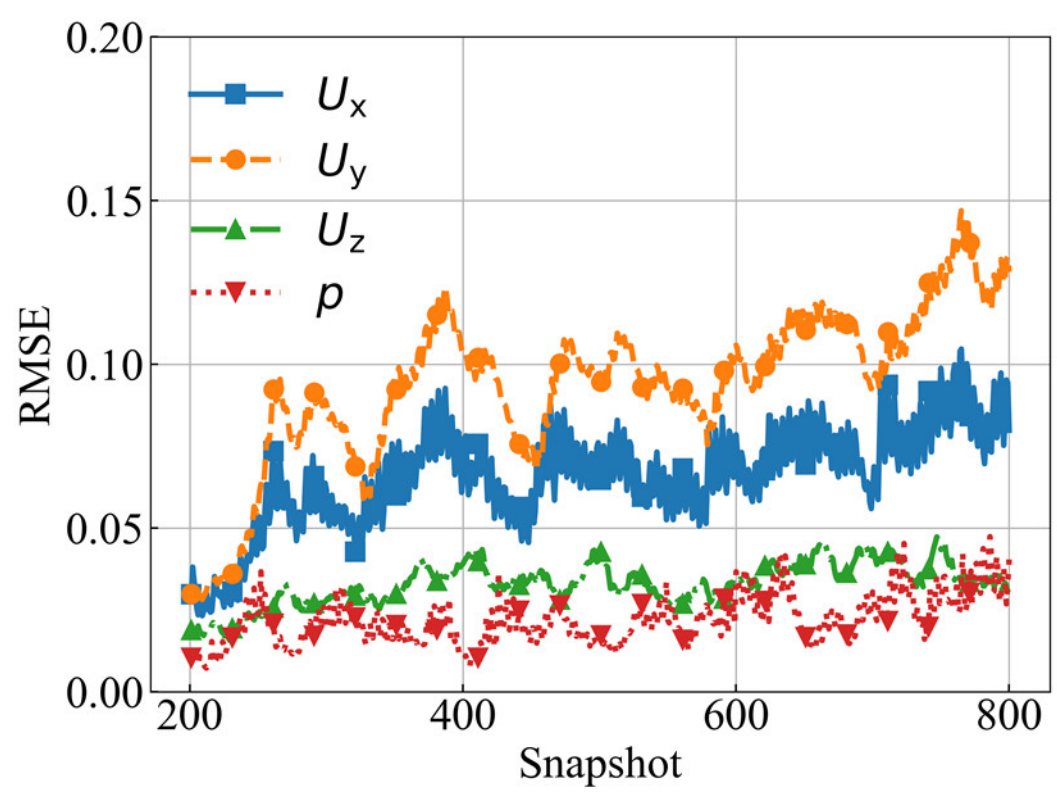

**Fig. 11.** Root mean square error (RMSE) between prediction and CFD ground truth for the flow past a circular cylinder at Re = 300.

### *4.2. Adaptive prediction with prescribed update intervals*

To mitigate the progressive degradation observed above, an adaptive strategy with prescribed update time intervals is first investigated. In this setting, snapshots (0, 200] with a sample time step of 0.5 s are used to train the initial surrogate, and the model predicts snapshots (200, 400]. The CFD solver is then resumed from the predicted field at snapshot 400 to compute snapshots (400, 600]. These newly generated snapshots are used to update the training dataset. The POD-DL model is subsequently updated using the enriched dataset, and a second prediction stage is performed for snapshots (600, 800]. In realistic applications of adaptive prediction, CFD ground truth is not assumed to be available during the prediction stage. However, for a posteriori evaluation only, CFD solutions over the two prediction intervals are computed separately and employed to validate the predicted results.

The case with Re = 300 is selected to assess this adaptive strategy. Compared with the Re = 200 case, the wake dynamics at Re = 300 are more complex and therefore more challenging for long-horizon autoregressive prediction, as discussed in Sec. 4.1. For this reason, in addition to instantaneous error metrics, it is necessary to examine whether the adaptive surrogate preserves the main statistical and temporal characteristics of the flow.

The first assessment is based on the turbulent kinetic energy (TKE) defined in Eq. (19). Fig. 12 compares the temporal evolution of the mean and standard deviation of TKE between the CFD reference (GT) and the adaptive prediction (Pred.). In both prediction intervals, the predicted signals remain within the range of the CFD reference and reproduce the overall fluctuation level with reasonable accuracy. In Fig. 12 (a), the mean TKE predicted by the adaptive surrogate follows the same order of magnitude and the same slowly varying envelope as the ground truth. Although local discrepancies increase as the prediction advances, the prediction captures the dominant level of resolved kinetic activity throughout both intervals. A similar conclusion can be drawn from Fig. 12 (b), where the standard deviation of TKE is also reproduced reasonably. The predicted standard deviation reflects the variability of the flow and follows the temporal modulation of the reference signal.

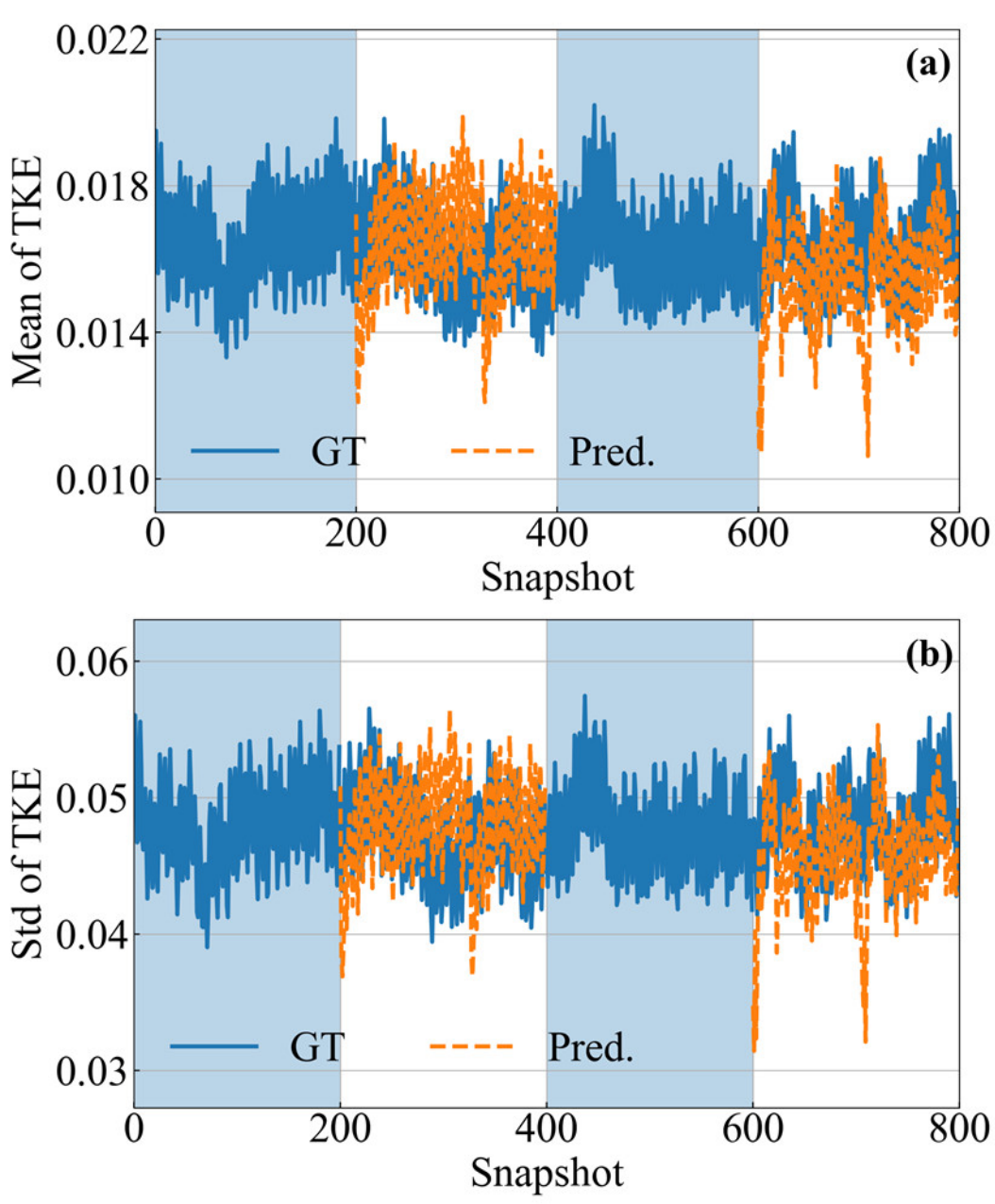

**Fig. 12.** Temporal evolution of the (a) mean and (b) standard deviation of turbulent kinetic energy for the CFD ground truth and the adaptive prediction.

To further evaluate whether the model retains the temporal organization of the wake dynamics, the temporal correlation of the streamwise velocity defined in Eq. (20) is examined at three probe locations, namely (0.5$D$, 0.5$D$, 0), (1$D$, 0.5$D$, 0), and (2$D$, 0.5$D$, 0), as shown in Fig. 13. As can be seen, the predicted temporal correlations exhibit a clear periodic oscillation in both prediction intervals, and the curves agree closely with the ground-truth results at the three locations. The positions of successive maxima and minima are well aligned, indicating that the surrogate preserves the dominant shedding frequency and the phase relationship of the wake. However, as the prediction advances within each interval, the agreement between prediction and ground truth gradually decreases, especially in terms of oscillation amplitude. This indicates that small autoregressive errors still accumulate over time, even though the dominant temporal periodicity is retained. Therefore, the adaptive model can recover the main temporal structure of the wake, but its instantaneous accuracy still decreases over extended forecast horizons.

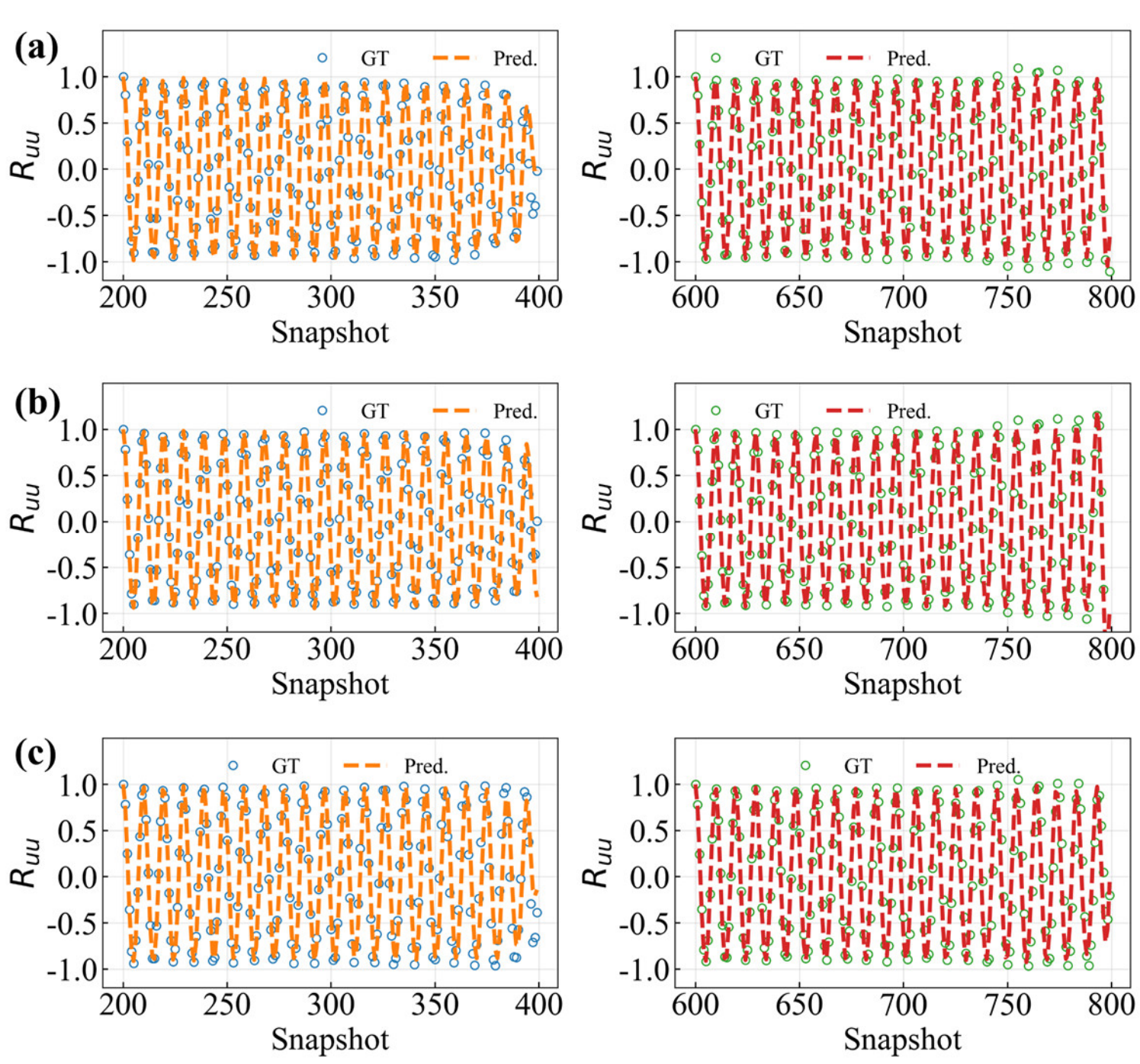

**Fig. 13.** Temporal correlation of streamwise velocity at probe locations (a) (0.5*D*, 0.5*D*, 0), (b) (1*D*, 0.5*D*, 0), (c) (2*D*, 0.5*D*, 0): comparison between CFD ground truth and adaptive prediction.

Agreement in statistical quantities alone is not sufficient to quantify the accumulated instantaneous error of the surrogate. For this reason, RMSE is computed for all predicted snapshots, and the corresponding results are shown in Fig. 14. For comparison, the figure also includes the RMSE obtained without adaptive prediction, which is plotted with semitransparent curves.

In both prediction intervals, the RMSE remains relatively low at the beginning and then gradually increases as the autoregressive prediction advances. This trend is consistent with the behavior discussed in Sec. 4.1 and reflects the cumulative nature of long-horizon forecasting errors. Even when the predicted flow remains statistically reasonable, small deviations may still accumulate over time and eventually produce larger instantaneous discrepancies.

In the second prediction stage, after the surrogate is updated using the CFD snapshots from 400–600, the RMSE for snapshots 600–800 is markedly lower than those associated with the later part of the first prediction interval. For example, the RMSE of $U_x$ and $U_y$ decreases by approximately 0.05 and 0.08 after retraining, respectively. In the non-adaptive reference, the errors continue to grow and remain at a relatively high level, whereas in the adaptive case the error is effectively reset after retraining and then evolves again from a substantially lower baseline. This behavior confirms the adaptive strategy improves the prediction.

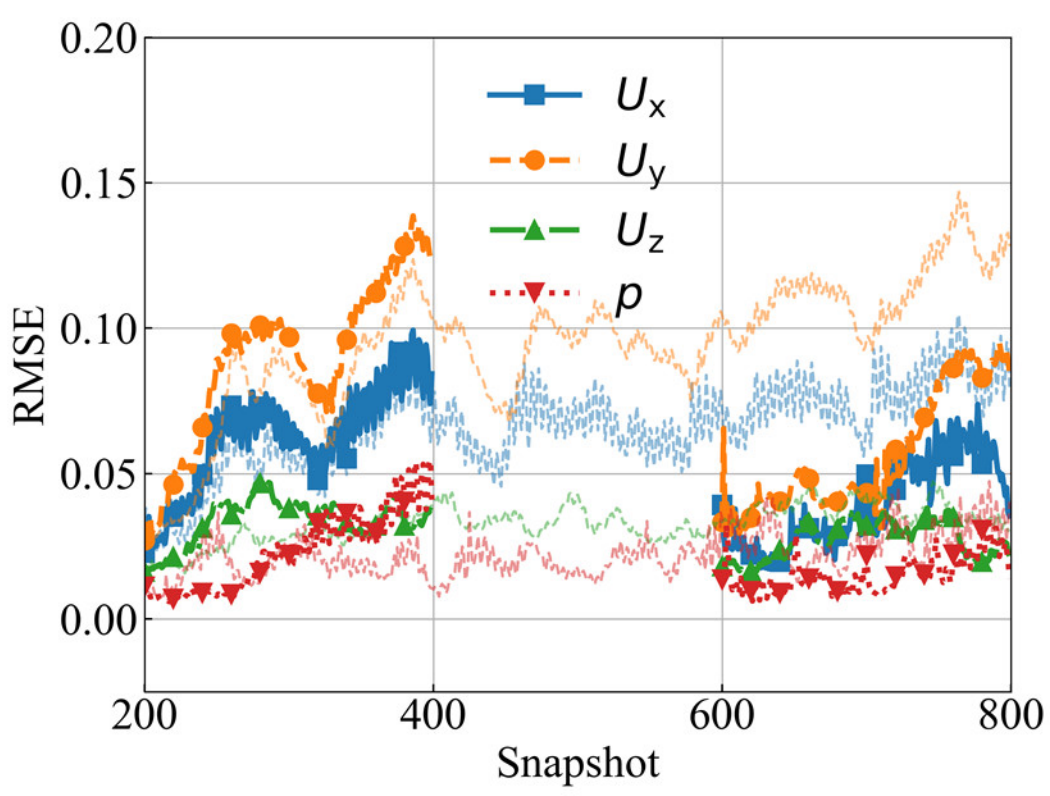

**Fig. 14.** Root mean square error (RMSE) between the prediction and the CFD ground truth: solid curves with markers denote adaptive predictions, whereas semitransparent curves without markers denote non-adaptive predictions.

In addition to prediction accuracy, the computational efficiency of the proposed adaptive framework is also evaluated. For a representative time interval of 200 snapshots, the computational time required by the CFD solver is 15906.9 s using 60 CPU cores, whereas the surrogate requires 2591.7 s using 4 CPU cores. When converted to single-core CPU time for a fair hardware-normalized comparison, the corresponding costs are 954414.0 s for CFD and 10366.8 s for the surrogate. The speed-up ratio, defined as the ratio between the CPU time required for CFD and that required by model training and prediction, is employed to evaluate acceleration performance. This gives an effective speed-up of approximately 92. Such a gain confirms that the adaptive framework can substantially reduce the computational burden while retaining satisfactory predictive fidelity.

Overall, the results in Figs. 12–14 demonstrate that adaptive prediction with a prescribed update interval provides an effective compromise between accuracy and computational cost. On the one hand, the surrogate preserves the main statistical properties of the wake, including the mean and fluctuation level of TKE and the temporal correlation of the streamwise velocity. On the other hand, the periodic reintroduction of CFD data prevents the irreversible growth of forecasting error and significantly improves long-horizon prediction accuracy relative to a purely offline, non-adaptive surrogate. In this sense, the prescribed-interval adaptive update acts as a controlled correction mechanism that repeatedly corrects the reduced-order predictor using high-fidelity information, thereby extending the practical prediction horizon while still avoiding continuous CFD computation.

*4.3. Event-triggered adaptive prediction by divergence detection*

The previous subsection considered adaptive updates at prescribed time intervals. Although effective, such a strategy may still resume the CFD solver earlier than necessary in some stages, or too late in others, because the update schedule is not directly linked to the actual reliability of the ongoing prediction. For this reason, a more flexible event-triggered adaptive strategy is introduced here, in which an indicator of divergence in the reduced space is monitored online and retraining is activated only when the surrogate begins to leave its trustworthy regime. In this way, the adaptive workflow responds directly to the evolving prediction state rather than a fixed external schedule.

In a realistic online setting, however, CFD results are not available during the prediction stage and therefore cannot be used as ground truth to evaluate error growth directly. This makes divergence detection intrinsically challenging. To address this issue, the present framework employs the ensemble-based uncertainty indicator introduced in Sec. 2.4. Specifically, three models trained with different random seeds are advanced simultaneously, and their prediction spread is used to quantify uncertainty. The corresponding threshold is updated dynamically from the recent uncertainty history. Once the uncertainty of a given variable exceeds its threshold for the prescribed criterion, the prediction for that variable is considered unreliable. Since all flow variables must remain temporally aligned in the returned prediction block, the final retained horizon is taken as the earliest trigger among all variables.

Fig. 15 shows the evolution of the ensemble uncertainty and the corresponding dynamic threshold during the first prediction stage. The raw ensemble uncertainty defined in Eq. (23) contains local short-time fluctuations. To make the trend clearer, it is smoothed using a moving average with edge padding. The smoothed curve suppresses isolated spikes and highlights the overall growth of uncertainty relative to the dynamic threshold. Meanwhile, the threshold varies smoothly as it is estimated from the recent uncertainty history. This behavior is important because the uncertainty magnitude is not constant throughout the prediction interval. A static threshold would either be too conservative in some stages or too permissive in others, whereas the present adaptive threshold defined in Eq. (28) is able to follow the evolving uncertainty baseline. In addition, the uncertainty evolution reveals variable-dependent reliability characteristics. For the velocity components, the uncertainty remains below the threshold over a relatively long interval before eventually rising toward the end of the prediction window. In contrast, the pressure prediction loses reliability earlier. According to Fig. 15, the uncertainty of $U_x$, $U_y$ and $U_z$ exceeds the threshold at approximately snapshot 325, whereas the pressure uncertainty crosses the threshold much earlier, at approximately snapshot 250. Since this prediction stage starts from snapshot 200, the pressure-triggered cutoff at approximately snapshot 250 leaves 50 accepted prediction snapshots. Because a consistent snapshot range must be returned for all variables, this earliest divergence event determines the retained horizon.

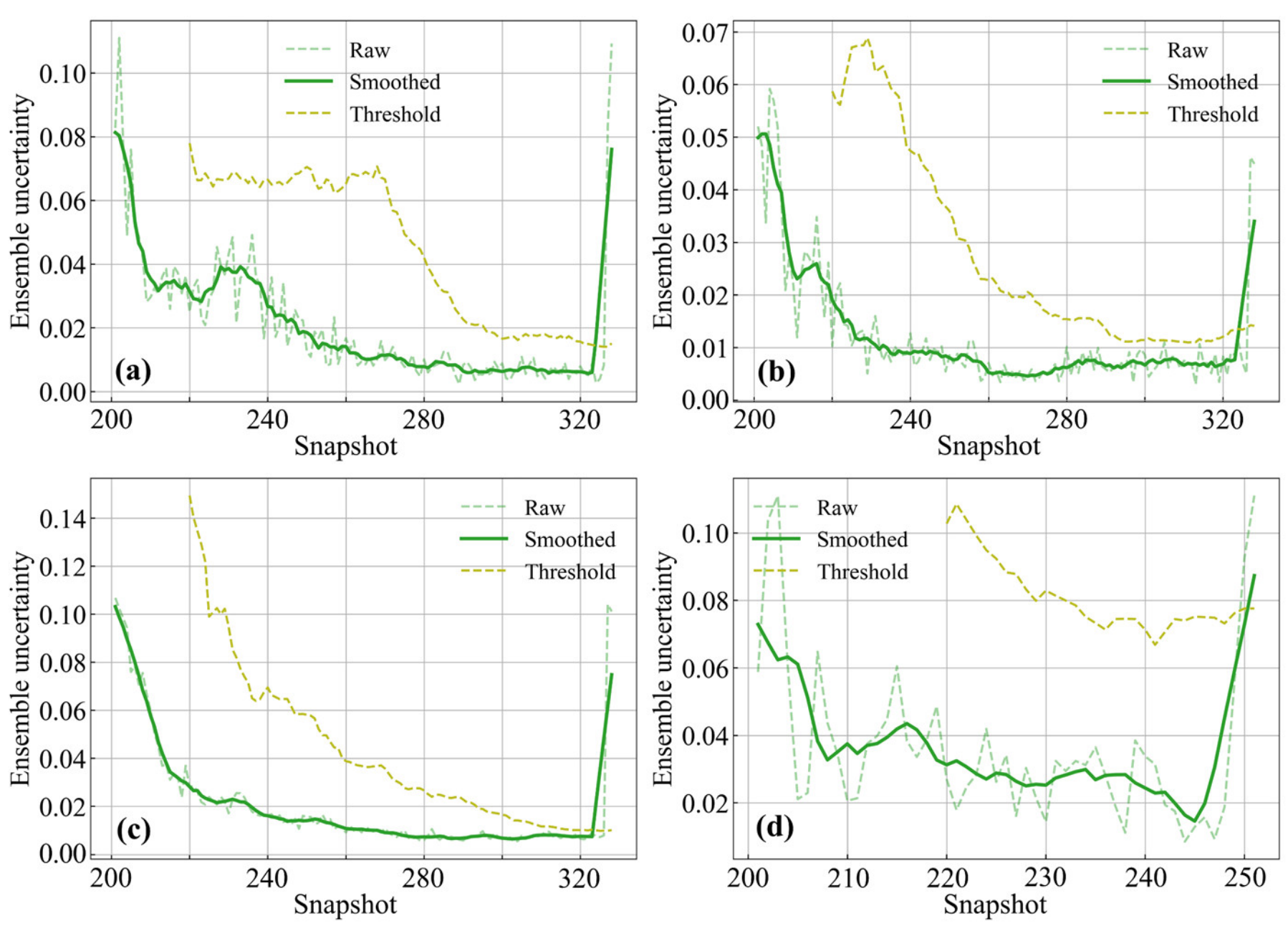

**Fig. 15.** Ensemble uncertainty and dynamically estimated threshold during the first event-triggered prediction stage for (a) $U_x$, (b) $U_y$, (c) $U_z$ and (d) $p$. Here, "Raw" denotes the original ensemble uncertainty, "Smoothed" the smoothed uncertainty, and "Threshold" the dynamically estimated threshold.

Once the first prediction stage is terminated, the CFD solver is invoked to generate new high-fidelity data from snapshot 250 to snapshot 450. These newly computed snapshots are used to update the surrogate model. The same event-triggered prediction is then repeated in the second stage. The corresponding uncertainty and threshold histories are shown in Fig. 16. Compared with the first stage, the second-stage prediction remains reliable over a slightly longer horizon. In particular, the uncertainty of $U_x$, $U_y$ and $p$ crosses the threshold at approximately snapshot 622, while that of $U_z$ crosses earlier, at approximately snapshot 560. As before, the returned horizon is determined by the earliest divergence among all variables. Thus, the second prediction stage is truncated according to the $U_z$ criterion. Even though the improvement in retained snapshots is moderate, the extension of the reliable horizon indicates that the updated surrogate has better adapted to the local flow regime after incorporating the new CFD information.

The differences between the first and second stages are also informative from a modeling perspective. In the first stage, pressure acts as the earliest indicator of divergence, whereas in the second stage the limiting variable becomes $U_z$. This change suggests that the dominant source of predictive uncertainty is not fixed, but depends on the flow region currently being traversed by the surrogate trajectory and on the specific information recently incorporated during retraining. In other words, the event-triggered framework does not merely detect that the prediction is degrading; it also provides variable-wise insight into how and where the reduced-order forecast starts to lose reliability.

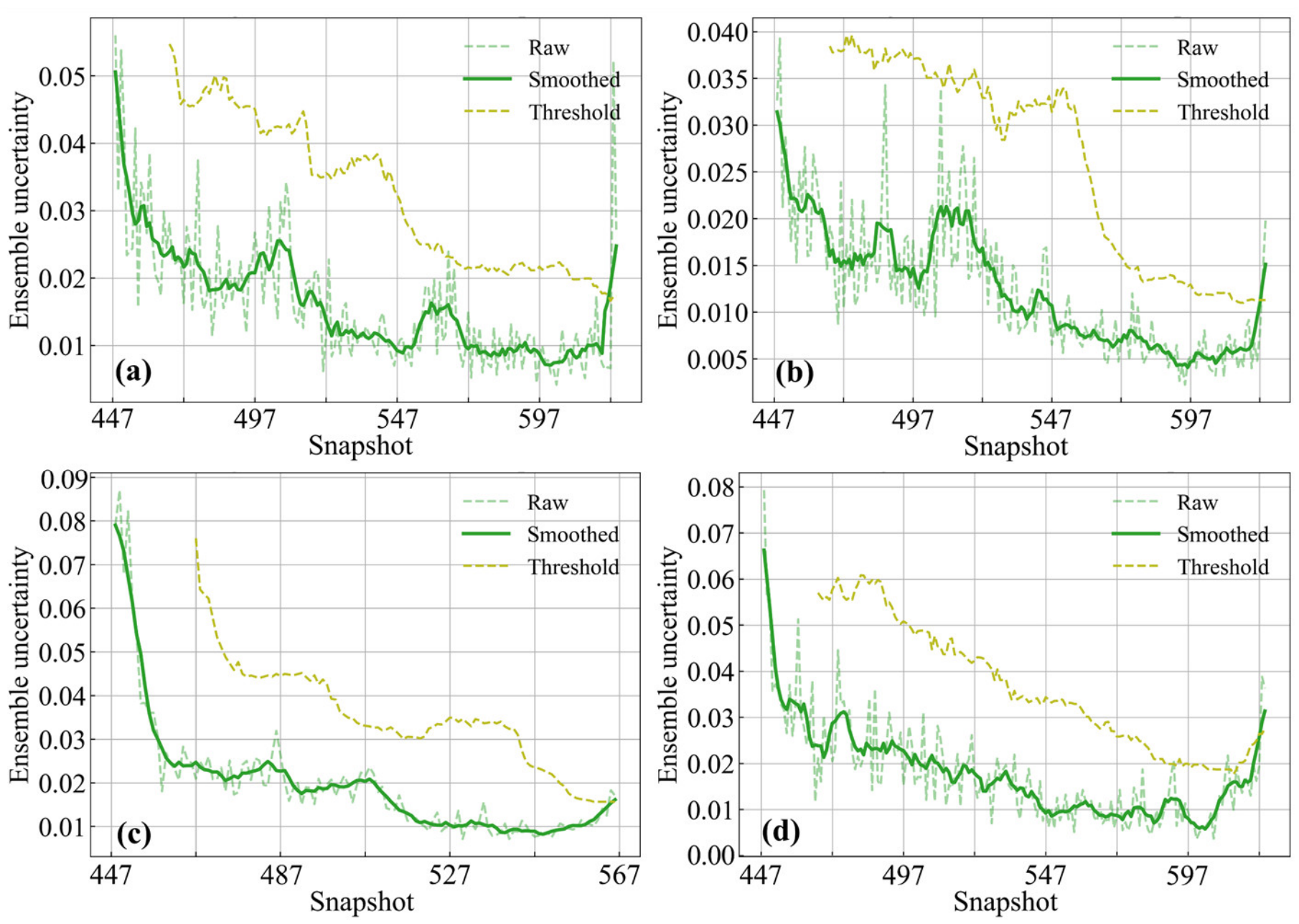

**Fig. 16.** Ensemble uncertainty and dynamically estimated threshold during the second event-triggered prediction stage after model updating for (a) $U_x$, (b) $U_y$, (c) $U_z$ and (d) $p$. Legend notation follows Fig.

15.

Since no CFD ground truth is available during the actual prediction stage, the quality of the event-triggered adaptive strategy is further examined through physically meaningful integral quantities, namely the drag coefficient $C_d$ and lift coefficient $C_l$. The original and smoothed histories of $C_d$ and $C_l$, obtained using Eq. (9), are shown in Fig. 17, together with the CFD-computed intervals indicated by the blue background.

For $C_d$, the predicted results reproduce the overall trend well in both stages. During the first stage, the predicted drag coefficient exhibits moderate fluctuations in the range 1.25–1.375, while in the second stage the oscillation level becomes smaller, with values mainly in the range of 1.25–1.34. It is also observed that the agreement with the slowly varying mean trend is improved. This indicates that the adaptive update helps stabilize the large-scale aerodynamic response of the wake. The lift coefficient $C_l$ provides an even more sensitive measure of predictive fidelity because it is strongly affected by the phase and symmetry of vortex shedding. As shown in Fig. 17 (b), the predicted $C_l$ reproduces the periodic oscillatory behavior of the CFD solution. More importantly, the locations where the prediction is truncated coincide with visibly deteriorating lift behavior. The two red dashed ellipses highlight stages at which the predicted $C_l$ begins to depart noticeably from the expected evolution. This observation supports the effectiveness of the uncertainty-based divergence detection: even though the CFD field is unavailable online, the ensemble uncertainty criterion is able to identify stages where the surrogate is about to lose physical fidelity. In this sense, the event-triggered mechanism is not merely a statistical alarm in reduced space, but is closely connected to degradation in physically interpretable output quantities.

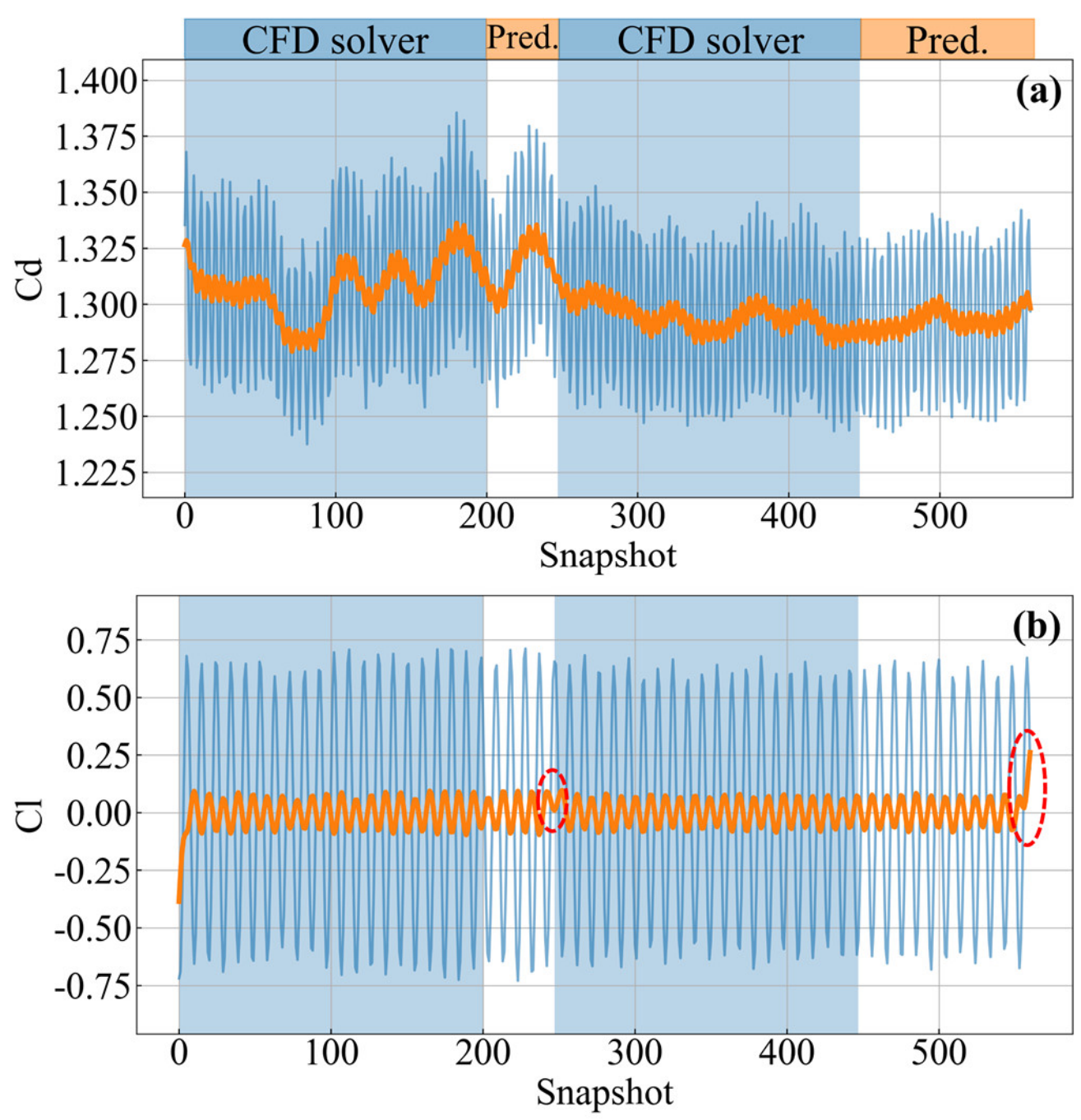

**Fig. 17.** Evolution of (a) drag coefficient $C_d$ and (b) lift coefficient $C_l$ during event-triggered adaptive prediction. Blue curves denote the original coefficients, and orange curves denote the smoothed coefficients obtained using the moving average defined in Eq. (9).

Overall, the results in Figs. 15-17 demonstrate that automatically adaptive prediction is more flexible than fixed-interval updating. By monitoring ensemble uncertainty and using dynamically estimated thresholds, the framework can stop the surrogate prediction when it becomes unreliable and recall CFD only when needed. This avoids both excessive CFD calls and delayed corrections. Although the accepted prediction length is still limited by the most sensitive variable, the method preserves the main evolution of the aerodynamic coefficients. It also provides a practical online strategy to balance robustness and efficiency when ground truth is not available.

*4.4. Adaptive prediction with varying input*

In this section, a further and more demanding test of the proposed framework is considered under varying operating conditions. Such situations are very common in realistic applications. When key parameters change, a surrogate model trained under previous conditions may no longer remain valid, and retraining becomes necessary. In practical systems, these changes can often be detected through sensors monitoring critical variables, which can then be supplied to the prediction framework as additional inputs. In this setting, the surrogate is no longer required merely to extrapolate in time along a single trajectory, but also to remain reliable when the input or control condition changes. This scenario is particularly relevant for practical CFD workflows, since many engineering systems operate under non-stationary boundary conditions and cannot be represented adequately by a single fixed-regime training set.

In the present study, the inlet velocity is selected as the varying input parameter. The prescribed inlet-velocity schedule during the CFD-prediction process is shown in Fig. 18 (a). The initial inlet velocity $U_{in}$ is 1 m/s. After the initial CFD simulation, 200 snapshots are generated and used to train the first surrogate model. The surrogate then starts prediction from snapshot 200. According to the imposed velocity schedule, the inlet velocity changes from 1 to 2 m/s at snapshot 300, then to 0.8 m/s at snapshot 600, and finally to 1.5 m/s at snapshot 900. The corresponding Reynolds number range is approximately 160–400. Therefore, this test examines whether the adaptive framework can detect changes in the inlet condition during prediction and update the surrogate accordingly.

Fig. 18 (b) shows the evolution of ensemble uncertainty during the first prediction stage. When the prediction passes snapshot 300, the prescribed inlet velocity changes from 1 to 2 m/s. The sudden change in operating condition causes the ensemble uncertainty to increase rapidly and exceed the dynamic threshold. This indicates that the prediction is leaving its reliable regime. The current prediction is then truncated, and the CFD solver is recalled from the last trusted state. During this recalled CFD stage, the inlet boundary condition is automatically updated to $U_{in}$ = 2 m/s, and 200 new high-fidelity snapshots are generated. These snapshots are employed as the training dataset and used to retrain the surrogate. A similar behavior is observed in Fig. 18 (c, d) for the subsequent inlet-velocity changes. In each case, the ensemble uncertainty rises after the operating condition changes, triggering CFD recall and model retraining. In this way, the framework can automatically detect regime changes, update the CFD boundary condition, generate new data, and adapt the surrogate to the new operating state.

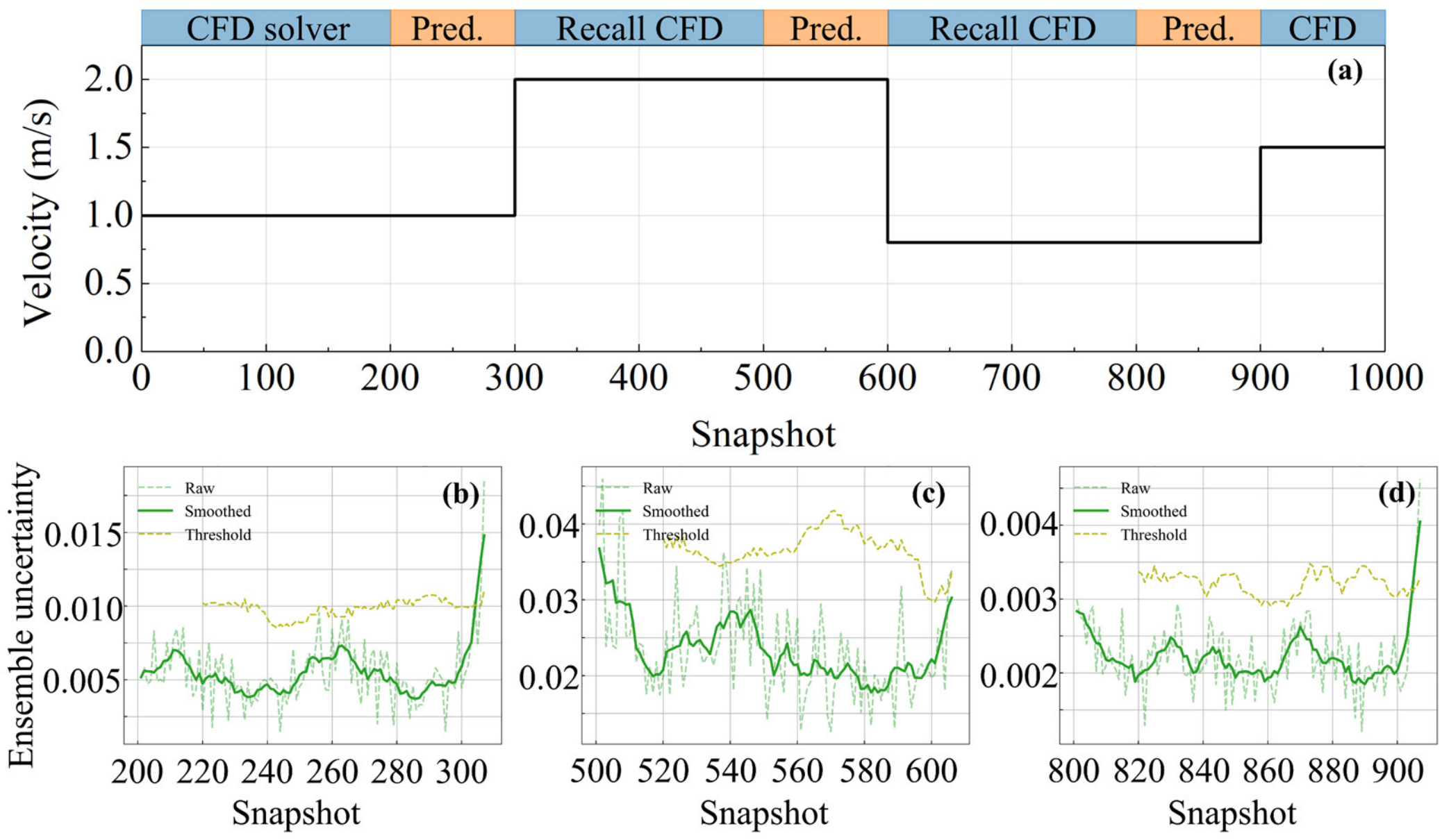

**Fig. 18.** (a) Variation of inlet velocity during CFD simulation/prediction, and the corresponding ensemble uncertainty of streamwise velocity during prediction for inlet-velocity stages of (b) 1 m/s, (c) 2 m/s and (d) 0.8 m/s.

To further assess the predictive performance, the evolutions of the drag and lift coefficients are compared in Fig. 19. As can be seen, abrupt changes appear in both figures around snapshots 300 and 600. These discontinuities occur because the CFD solver has just been recalled with a new inlet condition, while the flow field at that moment still reflects the prediction from the previous stage. Meanwhile, the reference velocity used in the computations of $C_d$ and $C_l$ is already updated to the new inlet velocity, which leads to an instantaneous jump in the force coefficients. As the CFD computation continues, both $C_d$ and $C_l$ gradually recover their physically expected oscillatory behavior, indicating that the wake has adjusted to the new operating condition.

Overall, the mean value of $C_d$ remains around 1.3, while $C_l$ oscillates around zero, as expected for the cylinder wake considered here. More importantly, the predicted results reproduce the main variation trend of the CFD solution. This is particularly evident in the evolution of $C_l$, whose oscillation frequency increases with increasing Re. As shown in Fig. 19 (b), the oscillation frequency increases when the inlet velocity changes from 1 m/s to 2 m/s, decreases when the inlet velocity is reduced to 0.8 m/s, and increases again when the inlet velocity rises to 1.5 m/s. This behavior indicates that the retrained surrogate captures the Re-dependent unsteady dynamics after each change in the inlet velocity. These results demonstrate that the present framework can identify input changes, recall the CFD solver, retrain the predictive model, and subsequently provide reliable forecasts under the updated inlet condition.

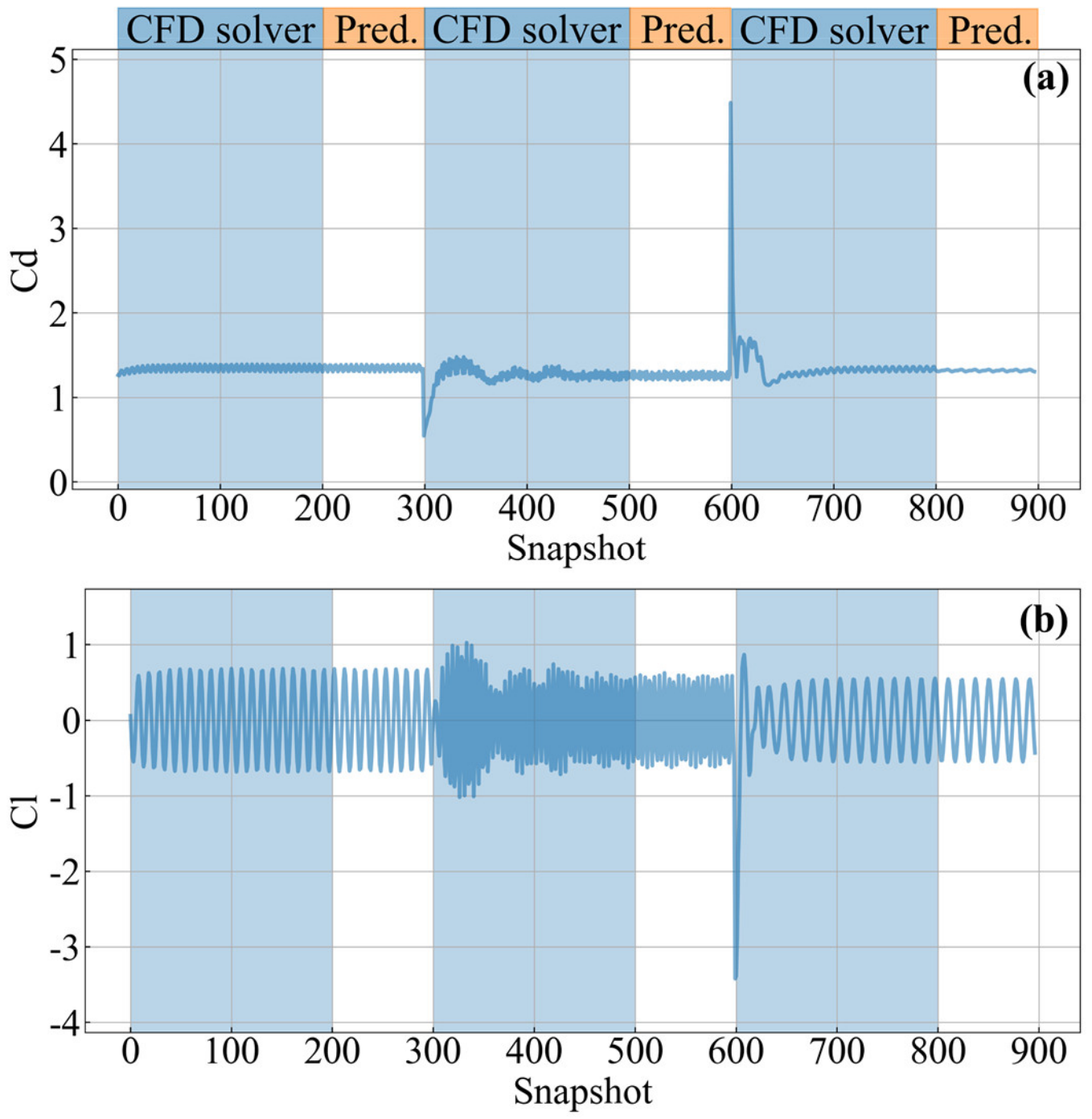

**Fig. 19.** Evolution of (a) drag coefficient $C_d$ and (b) lift coefficient $C_l$ under time-varying inlet velocity condition.

In summary, the present test confirms that the proposed adaptive framework is not limited to long-horizon prediction under fixed conditions, but can also operate effectively when the input varies with time. By combining online uncertainty monitoring, event-triggered CFD recall, and retraining, the framework is able to track regime changes and maintain predictive reliability across varying operating conditions. This capability is particularly important for realistic engineering applications, where operating parameters often vary during operation and purely offline-trained surrogates may quickly lose validity.

## 5. Conclusions

This work has presented a practical adaptive prediction framework that couples a CFD solver with a POD-DL model for long-horizon forecasting of unsteady flows in realistic online settings. Unlike conventional offline surrogates, which assume that the reference dynamics remain close to the training manifold, the present framework is designed to operate when the future flow state is unknown a priori. The central idea is to alternate between inexpensive surrogate forecasting and targeted high-fidelity solver recalls, so that the reduced-order model is continuously updated to the evolving flow dynamics.

The method is assessed on three-dimensional flow past a circular cylinder for Reynolds numbers in the range of 160–400. The baseline POD-DL surrogate is first examined without adaptation. Both testing cases, at Re = 200 and 300, show that prediction error increases gradually as the forecast horizon extends. A first adaptive strategy based on prescribed update intervals is then investigated. In this mode, the CFD solver is recalled after a fixed number of predicted snapshots, and the newly computed high-fidelity data are used as the training set to retrain the surrogate. The retrained model significantly reduces prediction errors during the second forecasting stage compared with the non-adaptive reference, highlighting the ability of the adaptive framework to mitigate long-term error accumulation. For a representative 200-snapshot interval, the framework achieves an effective speed-up of approximately 92 relative to CFD. The second strategy is an event-triggered adaptive mode. In this case, prediction reliability is monitored online through an ensemble-based uncertainty indicator defined in the POD space and combined with a dynamically estimated threshold. The results show that the uncertainty is able to identify the onset of unreliability before severe deterioration becomes apparent. Comparison with the evolution of drag and lift coefficients confirms that the event-triggered mechanism finishes prediction close to the onset of loss of physical fidelity, thereby providing an effective compromise between robustness and computational efficiency. Finally, the framework is tested under varying inlet velocity conditions, which represents a more realistic scenario in which the operating condition changes during numerical simulation. The inlet velocity is prescribed to model the variation of working condition. In this case, the ensemble uncertainty increases sharply when the inlet condition changes, triggering termination of the current surrogate forecast and recall of the CFD solver. After retraining with newly generated data under the updated condition, the surrogate recovers reliable predictive capability. The resulting drag and lift histories show that the adaptive framework captures the main changes in the cylinder wake.

The present results show that combining modal decompositions, deep learning, online uncertainty monitoring, and selective solver recall provides a viable route towards robust surrogates in practical workflows. The proposed adaptive method offers a practical framework for extending hybrid reduced-order modelling to complex unsteady flows in which long-term stability, regime transitions and changing operating conditions are critical.

**Acknowledgment**

The authors acknowledge the funding from the European Union's Horizon Europe research and innovation programme under the Marie Skłodowska-Curie grant agreements H2SAFIR No. 101269509, ENCODING No. 101072779, and MODELAIR No. 101072559. Views and opinions expressed are however those of the author(s) only and do not necessarily reflect those of the European Union or the European Research Executive Agency. Neither the European Union nor the granting authority can be held responsible for them. The authors also acknowledge the grant PID2023-147790OB-I00 funded by MCIU/AEI/10.13039/501100011033/FEDER, UE. The authors gratefully acknowledge the Universidad Politécnica de Madrid (www.upm.es) and CeSViMa for providing computing resources on the Magerit Supercomputer.

**Data availability statement**

The data that support the findings of this study are available from the corresponding author upon reasonable request.

**Appendix A. Pseudocode of adaptive prediction workflow**

In Sec. 2.1, the overall closed-loop adaptive prediction framework is described. For clarity, the pseudocode of the adaptive prediction workflow is summarized in Algorithm A.1.

**Algorithm A.1**

The pseudocode of the adaptive prediction workflow coupling OpenFOAM and POD–DL.

---

**Input**: OpenFOAM solver $S_{CFD}$, adaptive round index $n = 1$, initial simulation time $t_0$, final time $t_{final}$, snapshot interval $\Delta t_s$, prediction block length $N_{pred}$, CFD computational length $N_{cfd}$, POD truncation rank $r$, sequence length $q$, adaptation mode $M \in$ {scheduled, event-triggered}, burn-in length $N_{burn}$, consecutive trigger count $N_{consec}$, ensemble size $N_{ens}$.

---

**Output**: Predicted snapshots, updated dataset, updated POD-DL predictor.

**while** $t < t_{final}$ do

Run OpenFOAM from the latest accepted state and compute $N_{cfd}$ new snapshots.

Collect the snapshot dataset **A** from newly generated OpenFOAM data.

Mean-center the snapshot matrix, perform POD, and obtain POD coefficients $\mathbf{C} = \mathbf{V}_r^T$.

Train POD-DL predictor $f_\theta$ using **C**.

Construct the reduced initial condition from the most recent $q$ snapshots.

**if** $M$ = scheduled **then**

Perform block-wise autoregressive prediction with POD-DL for the next $N_{pred}$ steps.

Obtain predicted reduced coefficients $\hat{\mathbf{C}}$.

Decode $\hat{\mathbf{C}}$ to reconstruct the predicted fields $\hat{\mathbf{A}}$.

Accept the whole prediction block and write the predicted data in OpenFOAM format.

Set $t = t_0 + n(N_{cfd} \cdot \Delta t_s + N_{pred} \cdot \Delta t_s)$.

$n$ += 1.

**else if** $M$ = event-triggered then

Advance $N_{ens}$ surrogate models with different random seeds in parallel.

Compute the ensemble uncertainty $UC_{ens}(t)$ for each predicted step.

Update the dynamic threshold $\theta_\sigma(t)$ from the recent uncertainty history.

---

Ignore triggering during the burn-in interval $N_{\text{burn}}$.

Detect the earliest sustained exceedance satisfying $N_{\text{consec}}$ consecutive triggers.

**if** no trigger is detected within the current block **then**

Accept the whole prediction block and write the data in OpenFOAM format.

Set $t = t_0 + n * (N_{\text{cfd}} * \Delta t_s + N_{\text{pred}} * \Delta t_s)$.

$n$ += 1.

**else**

Determine the last reliable step $N_{\text{pred,trust}}$.

Accept only the reliable prediction sub-block up to $N_{\text{pred,trust}}$ and write data.

Set $t = t_0 + n * (N_{\text{cfd}} * \Delta t_s + N_{\text{pred,trust}} * \Delta t_s)$.

$n$ += 1.

**end if**

**end if**

**end while**

Obtain the whole flow fields and the updated predictor $f_\theta$.